\documentclass[fleqn,usenatbib]{mnras}

\usepackage{newtxtext,newtxmath}
\usepackage[T1]{fontenc}
\usepackage{ae,aecompl}

\usepackage{graphicx}	% Including figure files
\usepackage{amsmath}	% Advanced maths commands
\usepackage{amssymb}	% Extra maths symbols

\title[Roman galaxy redshift survey] {Clustering in the Simulated H$\alpha$ Galaxy Redshift Survey from Nancy Grace Roman Space Telescope}

\author[Z. Zhai et al.]{
Zhongxu Zhai,$^{1}$\thanks{E-mail: zhai@ipac.caltech.edu}
Chia-Hsun Chuang,$^{2}$
Yun Wang,$^{1}$
Andrew Benson,$^{3}$
Gustavo Yepes$^{4,5}$
\\
$^{1}$IPAC, California Institute of Technology, Mail Code 314-6, 1200 E. California Blvd., Pasadena, CA 91125 \\
$^{2}$Kavli Institute for Particle Astrophysics and Cosmology, Stanford University, 452 Lomita Mall, Stanford, CA 94305 \\
$^{3}$Carnegie Observatories, 813 Santa Barbara Street, Pasadena, CA 91101 \\
$^{4}$Departamento de F\'isica Te\'{o}rica, M\'{o}dulo 8, Facultad de Ciencias, Universidad Aut\'{o}noma de Madrid, 28049 Madrid, Spain \\
$^{5}$CIAFF, Facultad de Ciencias, Universidad Aut\'{o}noma de Madrid, 28049 Madrid, Spain \\
}
\date{Accepted XXX. Received YYY; in original form ZZZ}

\pubyear{2015}

\hypersetup{draft}
\begin{document}
\label{firstpage}
\pagerange{\pageref{firstpage}--\pageref{lastpage}}
\maketitle

\begin{abstract}

We present a realistic 2000 deg$^{2}$ H$\alpha$ galaxy mock catalog with $1<z<2$ for the Nancy Grace Roman Space Telescope galaxy redshift survey, the High Latitude Spectroscopic Survey (HLSS), created using Galacticus, a semi-analytical galaxy formation model, and high resolution cosmological N-body simulations. 
Galaxy clustering can probe dark energy and test gravity via baryon acoustic oscillation (BAO) and redshift space distortion (RSD) measurements.
Using our realistic mock as the simulated Roman HLSS data, and a covariance matrix computed using a large set of approximate mocks created using EZmock, we investigate the expected precision and accuracy of the BAO and RSD measurements using the same analysis techniques used in analyzing real data.
We find that the Roman H$\alpha$ galaxy survey alone can measure the angular diameter distance with 2\% uncertainty, the Hubble parameter with 3-6\% uncertainty, and the linear growth parameter with 7\% uncertainty, in each of four redshift bins. Our realistic forecast illustrates the power of the Roman galaxy survey in probing the nature of dark energy and testing gravity.
\end{abstract}


\section{Introduction}

Clustering of galaxies in the universe provides an important probe to constrain the relationship between galaxies and the underlying matter distribution. The accurate measurement can be used to describe the physics governing galaxy formation and evolution. In addition, the signal measured at different scales is also able to constrain the evolution of the large scale structure, including the linear growth rate through the redshift space distortion (RSD) effect and cosmic distance scales through the baryon acoustic oscillation (BAO). 

In the early universe, the photons and baryons are tightly correlated due to the high temperature. The pressure from the photons and gravitational interaction from the matter perturbations lead to oscillations in the coupled fluid. This feature is imprinted on the fluctuation of temperature and can be observed through cosmic microwave background (CMB), for instance the WMAP satellite (\citealt{Bennett_2003, Spergel_2003}) and Planck mission (\citealt{Planck_2014_1, Planck_2013}). This oscillating feature is also imprinted on the matter distribution, which can be measured through the power spectrum and correlation function of galaxies. Since the first detection of this BAO signal in the Sloan Digital Sky Survey (SDSS, \citealt{SDSS_York}) dataset (\citealt{Eisenstein_2005}) and the two-degree Field Survey (2dFS, \citealt{Cole_2005}), the technique of measuring and analyzing this phenomena has evolved significantly in the past decade. Based on observations from spectroscopic surveys like 6-degree Field Galaxy Survey (6dFGS, \citealt{Beutler_2011}), SDSS-II (\citealt{Ross_2015}), SDSS-III (\citealt{Dawson_BOSS, Alam_2016, Anderson_2014, Delubac_2015, Font-Ribera_2014}), SDSS-IV (\citealt{Dawson_2016, Bautista_2018, Ata_2018}), WiggleZ (\citealt{Blake_2011}), the BAO signal has been reported in different types of tracers, including the luminous red galaxies (LRG), emission line galaxies (ELG) and quasors (QSO). The redshift range is also not limited to local galaxies. The measurement from Lyman-$\alpha$ Forest is able to measure the BAO signal in auto- and cross-correlation at redshift up to $2.40$ (\citealt{Delubac_2015, Font-Ribera_2014}). The compilation of these state-of-the-art measurements of the BAO signal and the cosmic distance scales can place constraint on the fundamental cosmological parameters. Combination with other cosmological data such as CMB, Type Ia supernovae and so on can further constrain the cosmological model and its deviation from $\Lambda$CDM. This has been widely explored in the literature, see e.g. \cite{Wang_2007, Zhai_DE_2017, Zhai_2019JCAP} and references therein.

Due to peculiar velocities, the observed redshifts of the galaxies are different from the prediction for a homogeneous universe. This so-called Redshift Space Distortion (RSD) effect changes the observed galaxy density field and the inferred clustering signal (\citealt{Kaiser_1987}). However, this systematics also carries information of the universe and can be used to infer the linear growth rate of cosmic large-scale structure. With the galaxy catalog from 6dFGS (\citealt{Beutler_2012}), WiggleZ (\citealt{Blake_2011}), SDSS-II (\citealt{Chuang_2013a}), SDSS-III (\citealt{Beutler_2014}) etc, we can have measurements over a wide redshift range that provide complementary constraints on the cosmological model. This dynamic probe is different than the geometrical probe such as BAO and Type Ia SNe, and can distinguish dark energy and modified gravity models from the standard $\Lambda$CDM model when they have the same expansion history. 

As NASA's flagship survey for next generation, Nancy Grace Roman Space Telescope (hereafter Roman, \citealt{Green_2012, Dressler_2012, Spergel_2015}) will measure millions of emission line galaxies in its planned 5 year mission. The resulting data will provide uniform and powerful measurements of the BAO signal and the RSD effect, as well as information to study galaxy evolution and star formation histories. For future surveys like Roman and Euclid (\citealt{Laureijs_2011, Laureijs_2012}), we need to optimize the survey designs and evaluate their performance in constraining dark energy. In \cite{Merson_2018} and \cite{Zhai_2019MNRAS}, we estimate the number densities of the H$\alpha$ and [OIII] emission line galaxies that Roman and Euclid will be able to observe based on Semi-analytic model (SAM) of galaxy formation. This enables further investigation of the clustering properties of these galaxies. In this work, we move forward by making a realistic galaxy mock catalog of the data expected from the Roman galaxy redshift survey, the High Latitude Spectroscopic Survey (HLSS), and investigate the expected precision and accuracy of the BAO and RSD measurements from it using the same analysis techniques used in analyzing real data. The mock catalogs presented here can also be used to study other cosmic probes, such as clustering measurement at small scales, higher order statistics, cross correlation with different tracers of dark matter distribution and so on.

Note that the Fisher matrix based method is often used to forecast the performance of future surveys, but it underestimates the uncertainties by construction (\citealt{Tegmark_1997}).
In this paper, we apply the same methodology used in analyzing actual data to the simulated Roman galaxy survey data. 
We utilize the calibrated SAM to generate a synthetic Roman H$\alpha$ galaxy catalog.
This catalog has an area of $~2000$ deg$^{2}$ to be consistent with the current baseline mission design. The observed cosmic volume can contain millions of galaxies for clustering measurements. In order to estimate the uncertainty of the clustering measurements, we compute the covariance matrix by adopting a quick and economic method to generate a large number of approximate mocks (EZmocks) that have clustering consistent with the SAM derived mock (the simulated Roman data set).

Our paper is organized as follows: in Section 2, we introduce the method for producing the galaxy mocks from SAM and EZmocks. Section 3 describes the methodology for modeling the galaxy power spectrum, and Section 4 presents the results on the BAO/RSD measurements from the simulated Roman H$\alpha$ galaxy mock. We discuss and conclude in Section 5.

\section{Methodology for Producing Galaxy Mocks}

\begin{figure}
\begin{center}
\includegraphics[width=8.5cm]{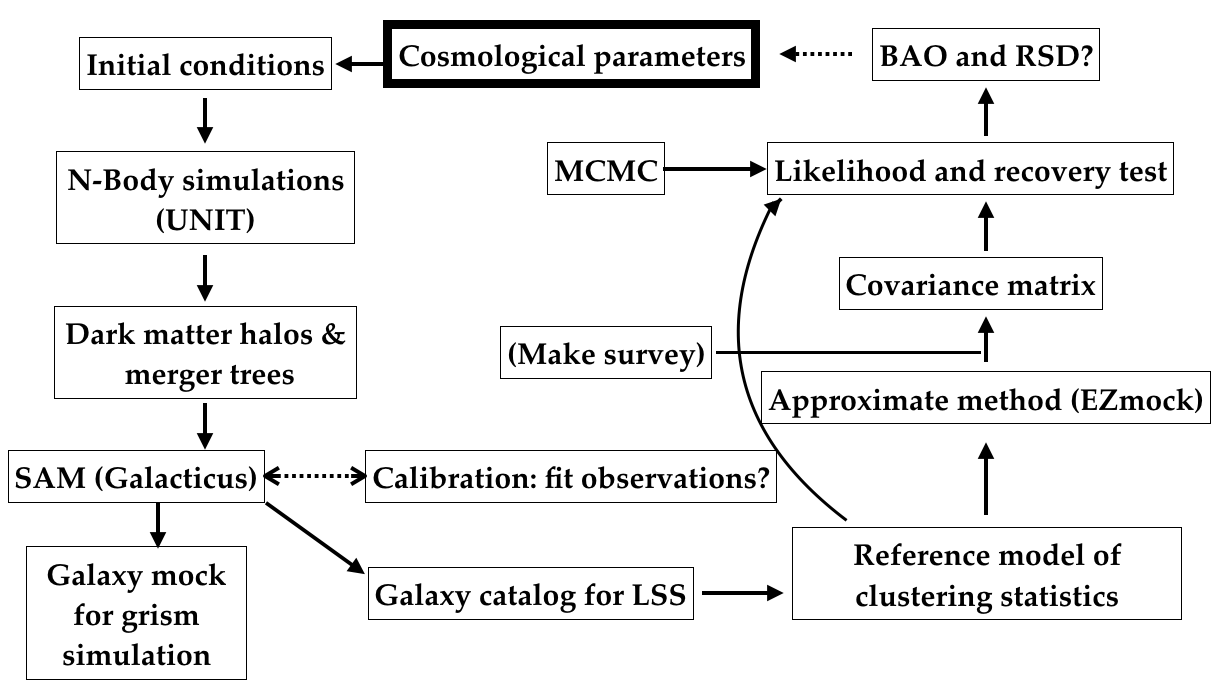}
\caption{Flowchart of the algorithm applied in this work, starting from a cosmological model. We assume a cosmology as applied in the UNIT simulation (\citealt{Chuang_2019}) throughout the work, i.e. the Planck 2016 cosmology (\citealt{Planck_2016}). With the dark matter halo and merger trees constructed from the N-body simulation, we calibrate the SAM to produce a galaxy catalog for the Roman galaxy survey. This catalog is assumed as the reference model to calibrate EZmock which is used to generate large number of approximate mocks for clustering analysis. We then estimate the covariance matrix for the clustering statistics of interest. This enables a recovery test to validate our simulation and modeling, as well as evaluate the precision and accuracy of BAO and RSD measurements from the Roman galaxy survey. Each step is explained in details in the text.}
\label{fig:flowchart}
\end{center}
\end{figure}

The analysis in this paper involves calibration and generation of multiple mocks for galaxy clustering, as well as the theoretical modeling to validate and interpret the result. Figure \ref{fig:flowchart} summarizes the whole algorithm in the analysis and each step is explained and described in following sections.

\subsection{Reference catalog from SAM}\label{sec:SAM}

The synthetic galaxy catalog can be constructed by populating high-resolution N-body simulations with galaxies. We adopt the SAM model based on \cite{Zhai_2019MNRAS}, which is used to predict the number counts of emission line galaxies that Roman will observe. In particular, we apply Galacticus (\citealt{Benson_2012}) to model galaxy formation and evolution. In \cite{Zhai_2019MNRAS}, this model is calibrated based on observations of the H$\alpha$ luminosity function from the ground-based narrow-band High-z Emission Line Survey (HiZELS, \citealt{Geach_2008, Sobral_2009, Sobral_2013}). In this work, we adopt the same parameters of Galacticus and the dust model to produce a reference catalog for clustering analysis. Note that \cite{Zhai_2019MNRAS} compare the prediction from two dust models, which can match the data from HiZELS or WISP number counts (\citealt{Mehta_2015}) respectively. For simplicity in this work, we just consider the dust model that can produce consistent result with HiZELS. The other dust model allows observation of more faint galaxies due to its weaker dust attenuation. We do not  expect it to significantly change the clustering analysis in this work. More details of the SAM can be found from \cite{Zhai_2019MNRAS}, and the dark matter halo merger trees used as input for the SAM calculation are from the UNIT simulation \footnote{\url{https://unitsims.ft.uam.es}} \citep{Chuang_2019}. We refer the readers to these references for more information and details. In addition, we also use the same SAM to produce a small simulated galaxy catalog along with spectral energy distributions as the input for the pixel level grism simulations for the Roman galaxy survey, to be presented in a separate paper.

\begin{figure}
\begin{center}
\includegraphics[width=9.5cm]{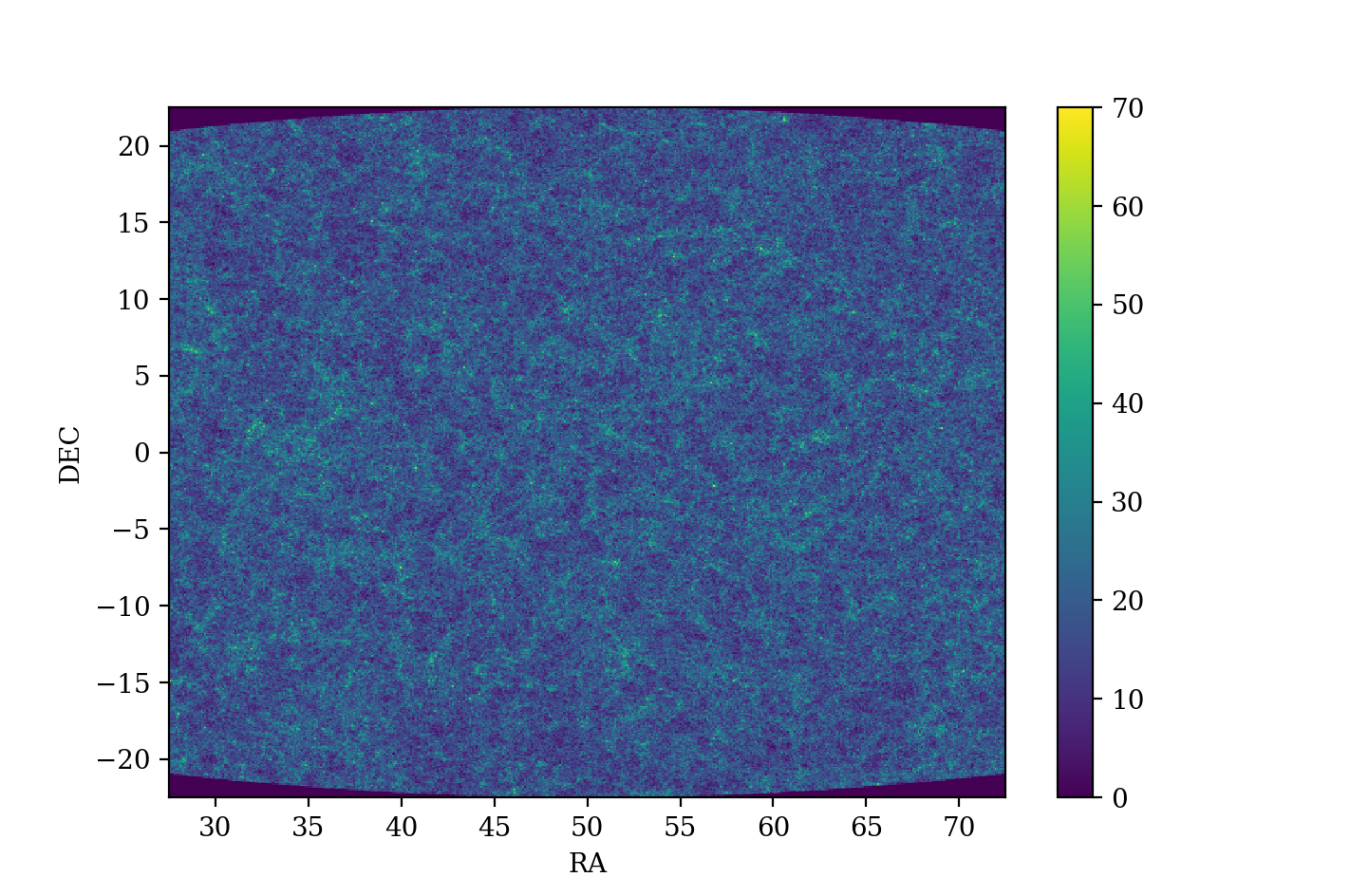}
\caption{The distribution of the mock galaxies generated by Galacticus for subsample at $1.0<z<1.2$, color coded by density on coordinates RA and DEC. Other redshift bins are omitted to avoid redundancy.}
\label{fig:RADEC}
\end{center}
\end{figure}

The Roman HLSS H$\alpha$ mock generated from Galacticus covers an area of $\sim$2,000 deg$^{2}$, consistent with the current baseline mission design. We select all the galaxies with dust-attenuated H$\alpha$ flux higher than $1\times10^{-16}$erg s$^{-1}$ cm$^{-2}$. In the redshift range $1.0<z<2.0$, the survey can observe 10 million H$\alpha$-emitting galaxies within a cosmic volume of 7(Gpc$/h)^{3}$. Figure \ref{fig:RADEC} displays the density distribution of galaxies at $1.0<z<1.2$ on the sky. The other redshift bins have the same RA/DEC distribution. For clustering analysis, we split the catalog into 5 redshift bins with width $\Delta z=0.2$. We first measure the two-point correlation function $\xi_{r}$ in configuration space by the LS estimator (\citealt{LS_1993})
\begin{equation}
    \xi_{r}=\frac{DD-2DR+RR}{RR},
\end{equation}
where DD, DR and RR are suitably normalized numbers of data-data, data-random and random-random pairs with separation $r$. The random catalog is generated to match the sky coverage as the data catalog i.e. have the same distribution for right ascension and declination and uniformly distributed on the sky. The redshifts of the random catalogs are random draws from the data such that both catalogs have consistent radial distributions. In order to obtain a stable measurement of clustering, the random catalog is 10 times larger than the data.

Due to peculiar velocities, the spectroscopically observed redshift of a galaxy is the combination of two components:
\begin{equation}
    z_{\text{spec}}=z_{\text{cosmo}}+\frac{v_{p}}{ac},
\end{equation}
where $z_{\text{cosmo}}$ is the redshift if the universe is homogeneous, $v_{p}$ is the peculiar velocity along the line-of-sight, $a$ is the cosmic scale factor with $a=1/(1+z_{\text{cosmo}})$ and $c$ is the speed of light. This second term leads to a distortion in the measured galaxy density field and thus the resultant clustering property, i.e. the RSD effect \citep{Kaiser_1987}. We add this effect to the galaxy catalog and remeasure the clustering. On large scales where the linear perturbation theory applies, the RSD effect increases the clustering amplitude compared to the expectation in real space and affects the significance of the BAO peak. 

The galaxy power spectrum is the two-point correlation function in Fourier space. For a given galaxy overdensity in Fourier space $\delta_{g}(\bold{k})$, the power spectrum is defined as 
\begin{equation}
    <\delta_{g}(\bold{k})\delta_{g}(\bold{k'})>=(2\pi^{3})P(k)\delta^{D}(\bold{k+k'}).
\end{equation}

For both correlation function and power spectrum, we can expand them in harmonics and represent the measurements as Legendre multipoles. For the galaxy correlation function, we can measure on a two-dimensional grid of separations of galaxies pairs perpendicular ($r_{p}$) and parallel ($\pi$) to the line-of-sight through
\begin{equation}
    \pi = \frac{\mathbf{s}\cdot\mathbf{l}}{|\mathbf{l}|}, \quad r_{p}=\mathbf{s}\cdot\mathbf{s}-\pi^2,
\end{equation}
where $\mathbf{l}=(\mathbf{s}_{1}+\mathbf{s}_{2})/2$, $\mathbf{s}=\mathbf{s_{1}-\mathbf{s_{2}}}$, $\mathbf{s_{1}}$ and $\mathbf{s_{2}}$ are the positions of a pair of galaxies (\citealt{Davis_1983, Fisher_1994}). The resultant Legendre multipoles are given as
\begin{equation}
    \xi_{\ell}(r)=\frac{2\ell+1}{2}\int_{-1}^{1}\xi(s, \mu)L_{\ell}(\mu)d\mu,
\end{equation}
where $L_{\ell}$ is the Legendre polynomial of order $\ell$, and $\mu=\pi/s$. The decomposition of the galaxy power spectrum can be done in a similar way.

\begin{figure*}
\begin{center}
\includegraphics[width=8.5cm]{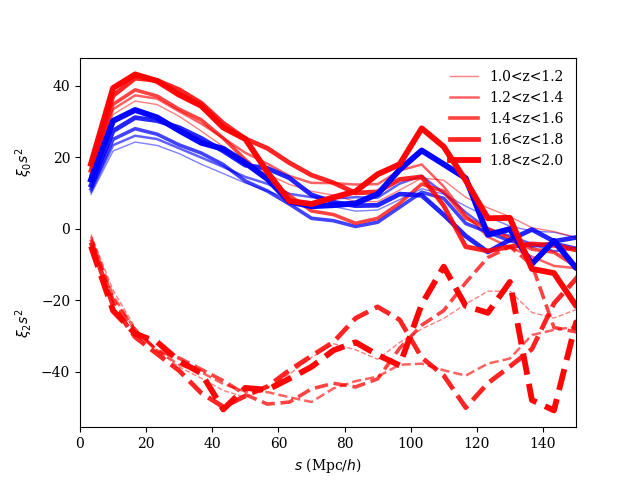}
\includegraphics[width=8.5cm]{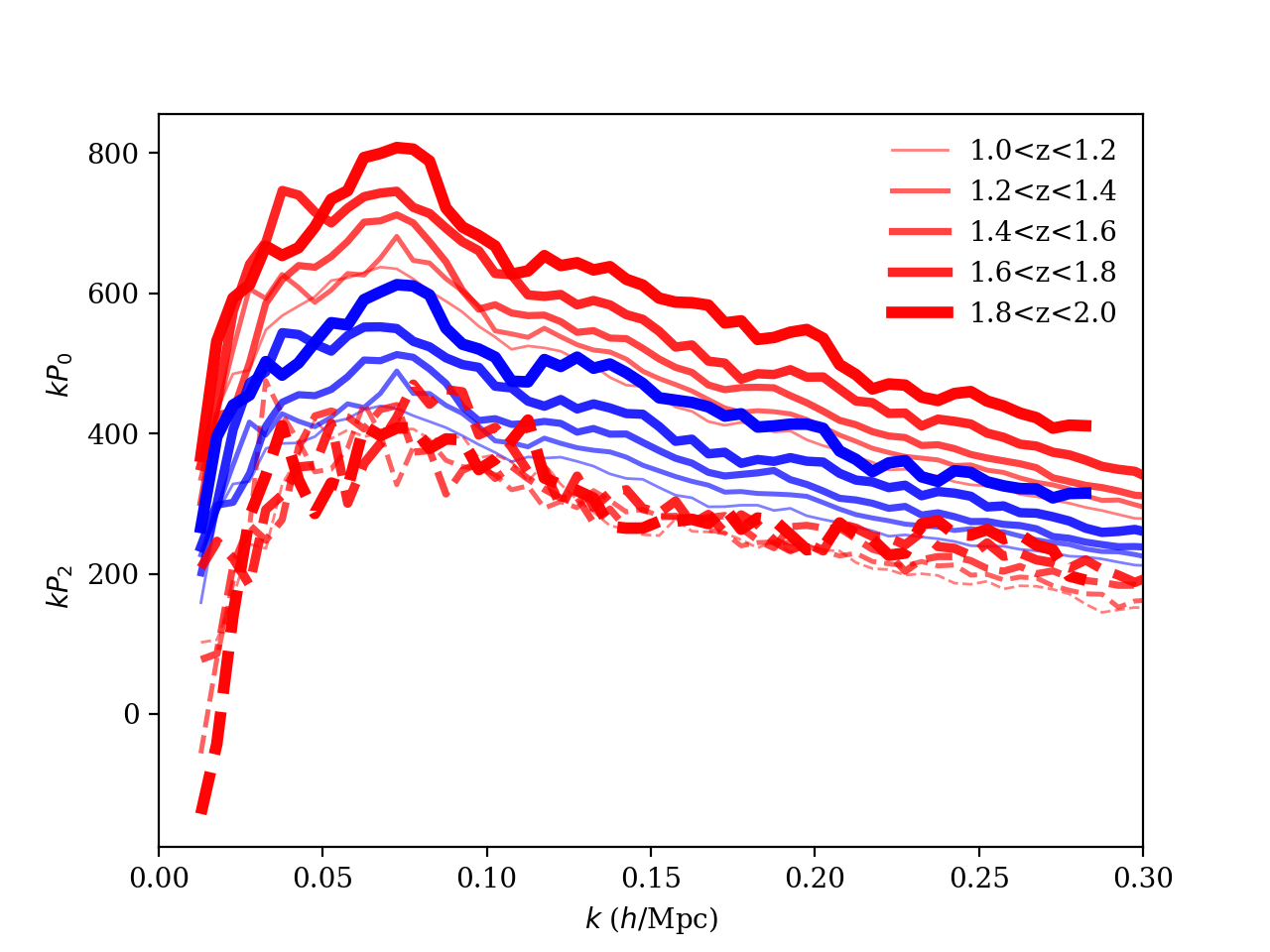}
\caption{Clustering measurements from the reference mock of Roman HLSS H$\alpha$ emission line galaxies generated by Galacticus. $Left$: 2-point correlation function, $Right:$ power spectrum. Solid lines show the monopole of the Legendre decomposition, while the dashed lines are the quadrupole. Blue is the result for real space, while red is in redshift space. Note that the Legendre multiples higher than monopole in real space is zero for both correlation function and power spectrum. Different line thickness denotes the redshift bins as shown in the legend.}
\label{fig:reference_clustering}
\end{center}
\end{figure*}

The clustering measurements of the galaxies in our simulated Roman HLSS H$\alpha$ mock are shown in Figure \ref{fig:reference_clustering}, for both correlation function and power spectrum in real and redshift space. 
We have used the publicly available code CUTE (\citealt{Alonso_2012}) to calculate the two-point correlation function, and Nbodykit (\citealt{Hand_2018}) to calculate the power spectrum. This mock catalog and its clustering measurement form the reference model in the construction of the large set of mock catalogs required for computing the covariance matrix.

\subsection{Approximate mocks from EZmock}

In order to assess the uncertainty of the measurements in a reliable manner, we need to estimate the covariance matrix of the measurements. The most robust approach is to employ a large number of mocks with clustering properties consistent with the data set. In principle, one can use the same SAM approach to generate the galaxy mocks. However, this requires a large number of N-body simulations to provide dark matter merger trees, which would be prohibitive in both cost and time using the current computational technology. Therefore we need an economic but accurate way to generate a sufficient number of galaxy mocks to estimate the covariance of the observable. This is the approach currently used in the analysis of actual data.

In this work, we use the EZmock code (\citealt{Chuang_2015}) to generate the approximate mocks. EZmocks are constructed using the Zel'dovich approximation of the density field and effectively include stochastic scale-dependent, non-local and non-linear biasing contributions. This approach has been demonstrated to be able to produce accurate clustering properties comparable to N-body simulations, including one-point, two-point and three-point statistics. This method has been widely used in the analysis of large scale structure, for instance in the BAO measurement from eBOSS QSO observations (\citealt{Ata_2018}).

EZmock has several parameters that need to be calibrated to fit the provided reference model. Since our reference catalog from SAM covers a wide redshift range of $1.0<z<2.0$, a single parameter set is not able to describe the evolution of the clustering. One solution is to measure the clustering in wider and overlapping redshift bins, measure the EZmock parameter values at different redshift and then interpolate to determine values at arbitrary redshifts where the clustering is measured, see for instance \cite{Ata_2018}. In this work, we calibrate the EZmock parameters at each redshift bins where the clustering signal is directly measured as shown in Section \ref{sec:SAM}.

In order to calibrate the EZmock parameters, i.e. find a parameter set for EZmock such that it can reproduce clustering signal as the reference model, we adopt an emulator-like method. Traditional interpolation method requires thousands or more sampling in the parameter space to achieve reasonable and accurate prediction at arbitrary point. The emulator method based on Gaussian Process (GP) or other machine learning algorithm can significantly reduce the required number of training data. We find that for modeling of the EZmock within a four-dimensional parameter space, a training set with less than one hundred data points is able to achieve acceptable accuracy. At a given redshift, we first generate a parameter set through a Latin hyper-cube algorithm uniformly distributed in the parameter space as the training set and use EZmock code to produce cubic mock for each model. In this calibration process, we use smaller box with boxsize 750 Mpc$/h$ and number of grid of 240 to accelerate the computation. Our test shows that it takes ~10 CPU minutes to produce one mock. Due to finite size and galaxy number density, the clustering measured from these mocks are noisy. We model the corresponding uncertainty by randomly choosing a training model and regenerate multiple mocks with different random seeds. We assume this can represent the level of error for the training set. The construction of the emulator follows the GP-based method as presented in \cite{Zhai_2019}. Since the dimension of the parameter space of the model is significantly lower than the modeling of galaxy clustering at small scale, a simpler kernel function in the GP construction is sufficient in this work. We choose the squared Exponential Covariance Function
\begin{equation}
    k_{\text{exp}}(r) = \exp\left(-\frac{r^{2}}{2l^{2}}\right),
\end{equation}
where the hyperparameter $l$ defines the characteristic length scale. We refer the readers to \cite{Zhai_2019} for more details in the construction of the emulator. 

\begin{figure*}
\begin{center}
\includegraphics[width=8.5cm]{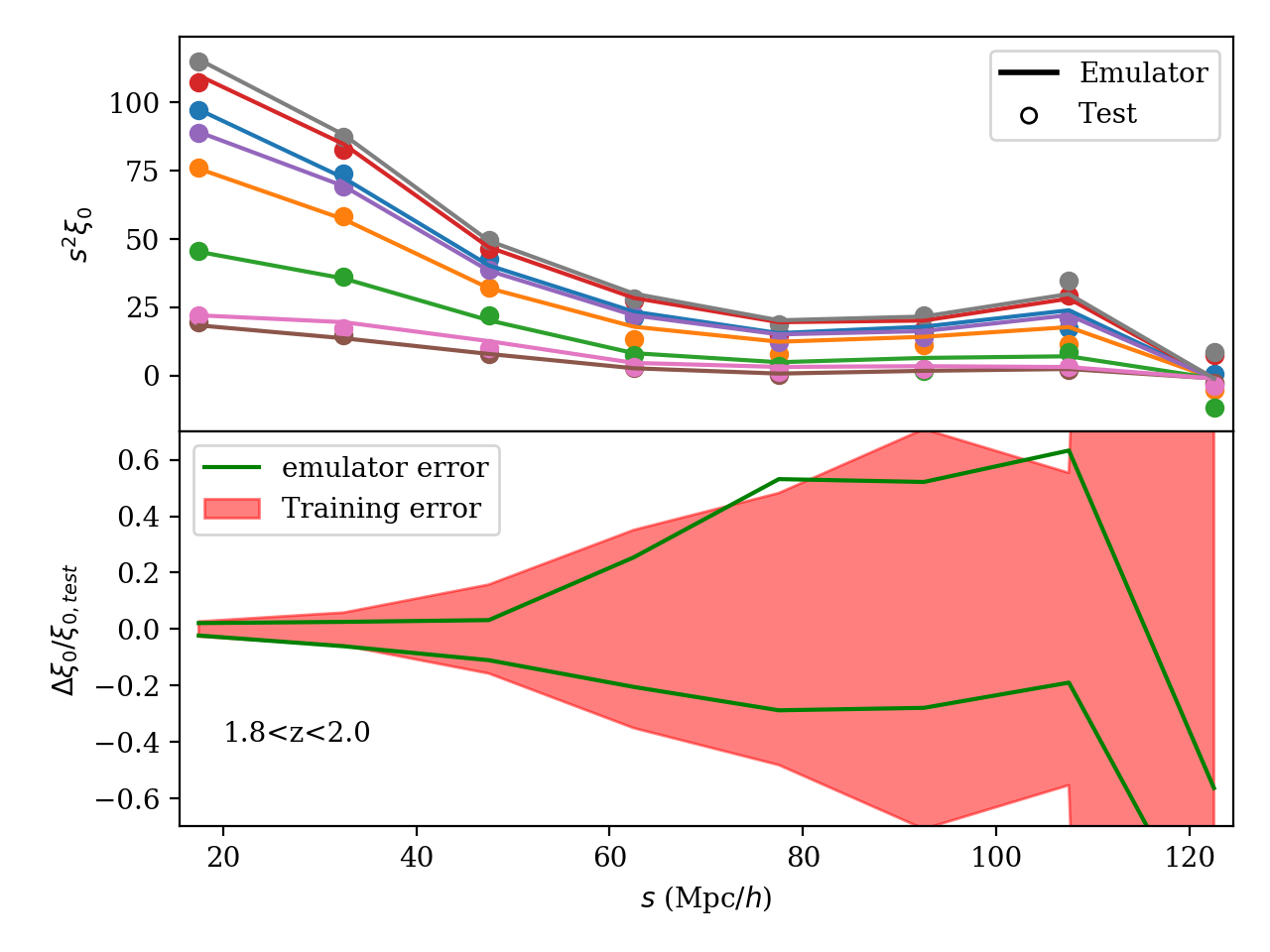}
\includegraphics[width=8.5cm]{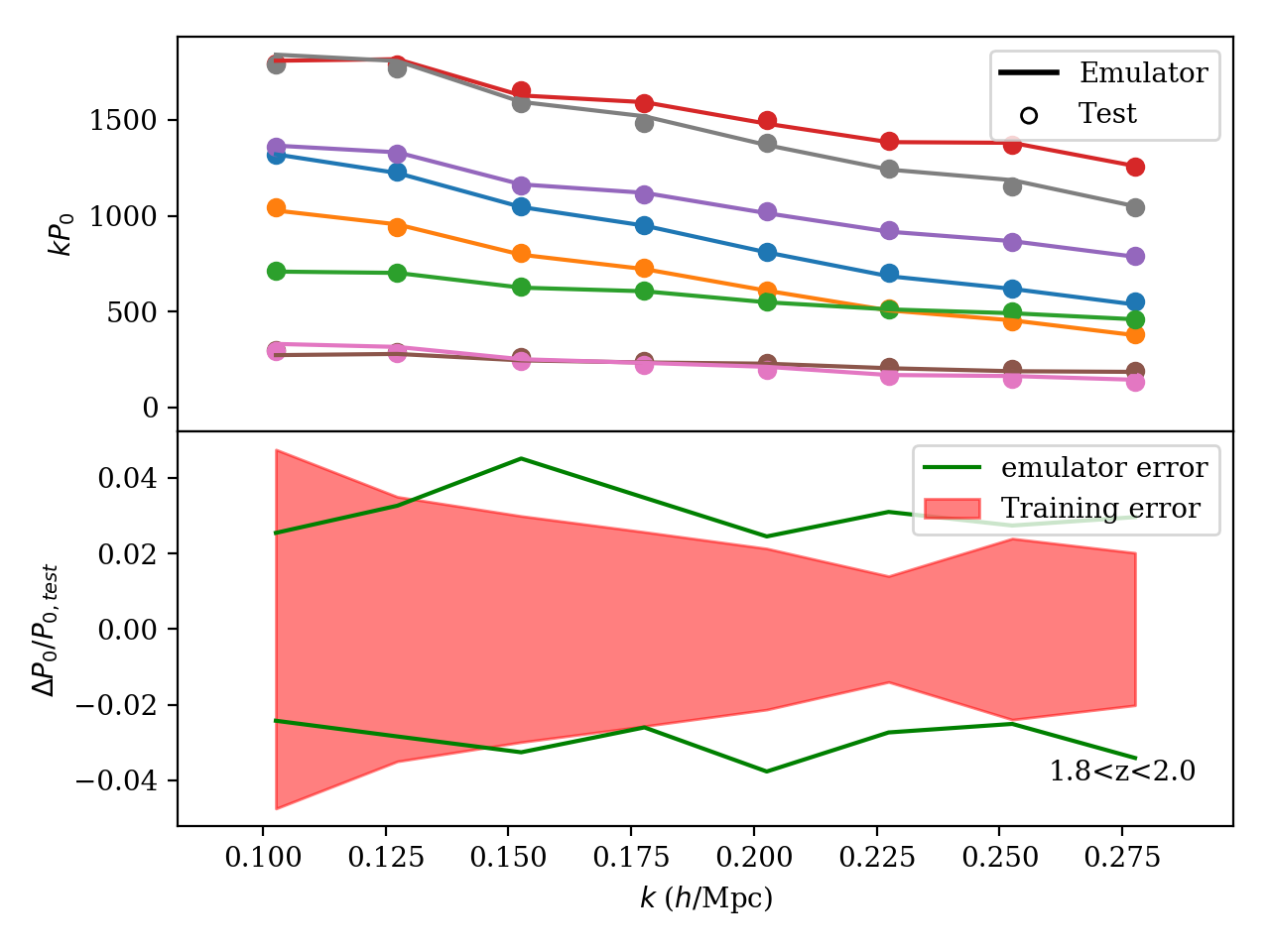}
\caption{Performance of the emulator built to calibrate the EZmock parameters. Only clustering statistics in redshift space for subsample within redshift $1.8<z<2.0$ is shown for illustration. The results for other redshift bins are similar due to the same algorithm. $Left:$ monopole of 2PCF, $Right:$ monopole of power spectrum. The top panel shows a few test models randomly chosen in the parameter space. Dots stand for the measurement from the EZmock directly, while the solid lines are the prediction of the emulator. The bottom panels show the overall performance of the emulator as the solid green line. The shaded area is 1$\sigma$ uncertainty estimated by multiple generations with different random seeds, also as the training error for the emulator. The emulator error is 68\% error estimated from 50 test models.}
\label{fig:emulator_accuracy}
\end{center}
\end{figure*}

In order to test the accuracy of the emulator, we also generate 50-100 tests points randomly distributed in the parameter space. Each model is also calculated by the EZmock code with the same boxsize and number of grid. In this work we focus on the two-point statistics of the mock galaxies. So we construct the emulators to predict the clustering statistics including the monopoles of correlation function and power spectrum in both real and redshift space. Note that in the analysis of actual data, only the redshift space is used in both calibration and interpretation. In particular, we summarize the following statistics for the emulator construction:
\begin{itemize}
    \item $\xi_{0}(s)$: monopole of 2PCF in both real and redshift space, with galaxy pair separation range of $20<s<120$ Mpc$/h$.
    \item $P_{0}(k)$: monopole of power spectrum in both real and redshift space, within range $0.1<k<0.3$ $h$Mpc$^{-1}$.
\end{itemize}
Adding higher Legendre multiples, like quadrupole and hexadecapole, or higher order statistics like bispectrum can improve the calibration of the EZmock parameter. We find that the clustering statistics described above are sufficient for the current purpose. In Figure {\ref{fig:emulator_accuracy}}, we show an example of the emulator performance for one redshift bin. The other redshift bins have similar results. In the top panel, we compare the results of the test sample which are directly calculated from EZmock as the truth, and the prediction from emulator. The overall error of the emulator is estimated as the 68\% distribution of the fractional error based on 50 test models. For 2PCF, we find that the emulator performance is better than the input training error at almost all the scales. The accuracy for the power spectrum is worse, however the uncertainty is better than 5\% at the scales of interest. We expect this error doesn't bias the calibration of the EZmock. In addition, we find that the error at large scale is significantly larger than intermediate and small scales, this is due to the sample variance of the finite volume, since the EZmock catalog constructed for calibration has a boxsize of 0.75 Gpc$/h$.

With the emulators to generate predictions of 2PCF and power spectrum for an arbitrary EZmock parameter set, we can search for the optimized model that can produce consistent result as the reference galaxy mock from SAM. This can be easily done through a $\chi^{2}$ computation
\begin{equation}
    \chi^{2}=(O_{\text{EZ}}-O_{\text{ref}})C^{-1}(O_{\text{EZ}}-O_{\text{ref}}),
\end{equation}
where $O_{\text{EZ, ref}}$ denotes the clustering statistics (2PCF or power spectrum) for EZmock prediction from emulator or reference model, $C$ is the corresponding covariance matrix. For simplicity, we just use the training error as for the emulator to get the covariance matrix here. We note that better estimate of this covariance matrix is possible by increasing the number of population and can improve the search for the best-fit parameters. However this is not of critical importance as the calibration is only used to match the reference model as close as possible.

Since the EZmock only gives cubic mock, we need to truncate and convert the mock to have the same radial and angular distribution as the reference mock. We use the  MAKE\_SURVEY toolkit (\citealt{White_2014}) to rotate and trim the cubic mock. In order to embed the whole survey volume of Roman HLSS for H$\alpha$ emitting galaxies in the simulation, we produce EZmock catalog with boxsize of 3 Gpc$/h$ and number of grid of 960. Using the code MAKE\_SURVEY, we construct the mock containing the whole survey range within redshift $1<z<2$ without re-using any region of the box.
For each subsample of the clustering mock, we generate 1000 EZmocks which is sufficient to estimate the covariance matrix for the clustering statistics.

\begin{figure*}
\begin{center}
\includegraphics[width=18.5cm]{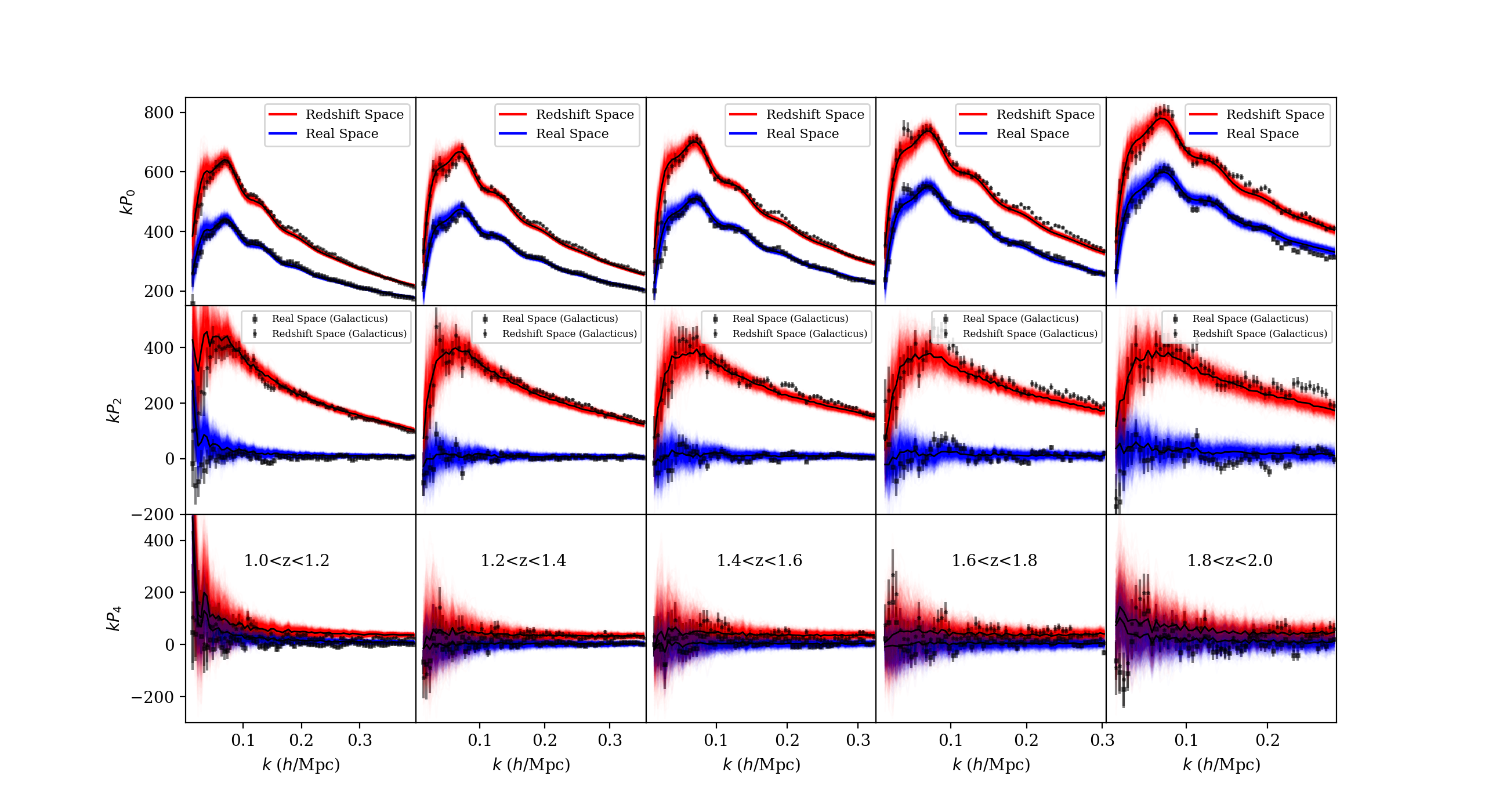}
\caption{The power spectrum measured from the reference catalog and calibrated EZmocks. The EZmocks are generated with boxsize=3Gpc$/h$ to embed the whole Roman HLSS with $1<z<2$, and further truncated to have the same angular and radial distribution as the reference model. The results for different redshift bins are shown in each column. The top row shows monopole of galaxy power spectrum, the middle row shows quadrupole and bottom row shows hexadecapole. The red lines are measurement from EZmocks in redshift space and blue lines are in real space. Note that the quadrupole and hexadecapole in real space are zero, which are also shown for completeness. The solid lines are the mean of the 1000 EZmocks and the error bars in the measurement of reference model are uncertainties estimated from EZmocks. The calibration of the EZmock only utilizes information from monopole, the consistency of all the three multipoles with the reference model shown here validates the modeling for the galaxy clustering.}
\label{fig:pk_ezmock}
\end{center}
\end{figure*}

In Figure \ref{fig:pk_ezmock}, we present the final measurement of the power spectrum from the calibrated EZmocks, which is truncated to have the same angular and radial distribution as the reference catalog from Galacticus. The result shows the measurement of monopole, quadrupole and hexadecapole in both real and redshift space. Note that the calibration of EZmock only uses the monopole, the consistency of higher Legendre multipoles between EZmock and Galacticus validates our calibration and modeling of these approximate mocks. 

\section{Galaxy clustering modeling}

\subsection{Power spectrum model}\label{sec:pk}

The Roman HLSS H$\alpha$ mock galaxy catalog and its associated approximate galaxy mocks we have built enable the validation of the algorithm for extracting cosmological parameters. For investigating BAO/RSD measurements, we use a theoretical power spectrum to model the various known effects and compare with data, in deriving the measurement uncertainties on the cosmic distance scales and linear growth rate of large scale structure. We model the galaxy power spectrum as \cite{Chuang_2013b, wang_2014, Hemantha_2014}
\begin{equation}\label{eq:power1}
    P_{\text{dw}}^{\text{S}}(k, \mu, z)=P_{\text{dw}}(k, \mu, z)\frac{(1+\beta\mu^{2})^2}{1+\frac{1}{2}(k\mu\sigma_{v})^{2}},
\end{equation}
where $\beta$ is the linear redshift distortion parameter, $\sigma_{v}$ is the pairwise peculiar velocity (\citealt{Hamilton_1998}). $P_{\text{dw}}(k, \mu, z)$ is the dewiggled linear galaxy power spectrum given by
\begin{equation}
    P_{\text{dw}}=G^{2}P_{0}k^{n_{s}}T_{\text{dw}}(k, \mu, z),
\end{equation}
where $G(z)$ is the linear growth factor at redshift $z$, and $n_{s}$ is the power-law index of the primordial power spectrum. In order to construct the de-wiggled transfer function, we apply the method as in \cite{Wang_2013_Tk}
\begin{eqnarray}
    T_{\text{dw}}(k,\mu, z)=T_{\text{lin}}^{2}(k,z)\exp\{-g_{\mu}k^{2}/k_{*}^{2}\}  \\
+T_{\text{nw}}^{2}(k,z)(1-\exp\{-g_{\mu}k^{2}/k_{*}^2\}),
\end{eqnarray}
where the linear transfer function $T_{\text{lin}}$ is calculated using CAMB \citep{Lewis_2000}, the ``no wiggle" transfer function $T_{\text{nw}}(k,z)$ is from equation (29) of \cite{Eisenstein_1998}, $k_{*}$ represents the scale of the non-linear effect on the baryon acoustic oscillation scales, and $g_{\mu}$ describes the enhanced damping along the line of sight due to the enhanced power given by
\begin{equation}
\label{eq:g_mu}
    g_{\mu} = G^{2}(z)\{1-\mu^{2}+\mu^{2}[1+f_{g}^*(z)]^{2}\},
\end{equation}
where $f_{g}^*$ is related to the linear growth rate.

Due to the non-linear evolution of the galaxy power spectrum, we use the model introduced by \cite{Cole_2005} to correct the linear matter power spectrum. The resultant galaxy power spectrum can be written as
\begin{equation}
    P_{\text{nl}}^{S}(k, \mu, z)=\frac{1+Qk^{2}}{1+Ak+Bk^{2}}P_{\text{dw}}^{S}(k, \mu, z),
\end{equation}
where $A, B$ and $Q$ are constants. \cite{Sanchez_2008} fix $B=Q/10$ in the modeling of the non-linear power spectrum at small scales. In the following analysis in this work, we also allow $B$ to be a free parameter to make the non-linear correction more flexible.

We note that the RSD model for the galaxy power spectrum, Eq.[\ref{eq:power1}], is not unique in the analysis of galaxy clustering, and different models have been utilized in literature. In this work, we also adopt a different model to compare the results. In particular, we compare a model as demonstrated in \cite{Wang_2017}:
\begin{equation}\label{eq:power2}
    P_{\text{dw}}^{\text{S}}(k, \mu, z)=P_{\text{dw}}(k, \mu, z)\left(1+\beta \bar{W}(k,z)\mu^{2}\right)^2\exp\left[-\frac{(k\mu\sigma_{v})^{2}}{2}\right],
\end{equation}
where the window function takes the form (\citealt{Zheng_2013, Zhang_2013})
\begin{equation}
    \bar{W}(k,z)=\frac{1}{1+\Delta\alpha(z)\Delta^{2}(k,z)},
\end{equation}
where $\Delta\alpha(z)$ is a parameter to be determined by observational data, $\Delta^{2}(k,z)=k^{3}P_{\text{lin}}/(2\pi^2)$, and $P_{\text{lin}}$ is the linear matter power spectrum. Compared with the model in Eq.[\ref{eq:power1}], this model changes the RSD modeling in the galaxy clustering, and the galaxy peculiar velocity distribution is also changed. We denote Eq.[\ref{eq:power1}] as Model A and Eq.[\ref{eq:power2}] as Model B in our analysis. In Model A, the galaxy peculiar velocity distribution takes the usual form $f(v)=(\sigma_{v}\sqrt{2})^{-1}\exp(-\sqrt{2}|v|/\sigma_{v})$ which in Fourier space corresponds to one over the denominator in Eq.[\ref{eq:power1}]. In Model B, the galaxy peculiar velocity distribution takes the Gaussian form $f(v)=1/(\sigma_{v}\sqrt{2\pi})\exp(-v^2/(2\sigma_{v}^2))$.

In the actual measurement of the galaxy power spectrum, we need to assume a cosmological model to convert galaxy's position from angular and redshift coordinates into 3D comoving coordinates. Therefore a reference model or fiducial model is applied. However, it is possible that the fiducial model is different than the true cosmology, this can lead to a distortion in the observed galaxy power spectrum than the true power spectrum, known as AP effect (\citealt{Alcock_1979}). In order to take into account this effect in the modeling of the galaxy power spectrum, one can introduce two scaling parameters 
\begin{equation}
    \alpha_{||}=\frac{H^{\text{fid}}(z)r_{s}^{\text{fid}}(z_{d})}{H(z)r_{s}(z_{d})}, \quad
    \alpha_{\perp}=\frac{D_{A}(z)r_{s}^{\text{fid}}(z_{d})}{D_{A}^{\text{fid}}(z)r_{s}(z_{d})},
\end{equation}
where $H^{\text{fid}}(z)$ and $D_{A}^{\text{fid}}$ are the Hubble parameter and angular diameter distance for the fiducial model, and $r_{s}^{\text{fid}}(z_{d})$ is prediction from the fiducial model for the sound horizon at the drag epoch. These two scaling parameters can relate the true wave vector component $k_{||}'$ and $k_{\perp}'$ with the observed ones: $k_{||}'=k_{||}/\alpha_{||}$ and $k_{\perp}'=k_{\perp}/\alpha_{\perp}$. This can further transfer to the coordinates of $k$ and $\mu$ via
\begin{eqnarray}
    k'=\frac{k}{\alpha_{\perp}}\left[1+\mu^{2}\left(\frac{1}{F^{2}}-1\right)\right]^{1/2} \\
    \mu'=\frac{\mu}{F}\left[1+\mu^{2}\left(\frac{1}{F^{2}}-1\right)\right]^{-1/2},
\end{eqnarray}
where $F=\alpha_{||}/\alpha_{\perp}$. Note that the AP effect can also change the overall amplitude of the power spectrum multipoles due to the volume difference compared with the fiducial model, which is degenerate with the amplitude of the power spectrum template $P_{0}$.

Following \cite{Wang_2013_Tk}, we model the constraints on the growth of large scale structure by measuring the {\it linear growth parameter}
\begin{equation}
\label{eq:fgsigmam}
    f_g(z) \sigma_m(z) \equiv f_g(z) G(z) \sqrt{P_0},
\end{equation}
where $P_0$ is the linear matter power spectrum normalization. The advantage of using $f_g(z) \sigma_m(z)$ over $f_g(z) \sigma_8(z)$ is that it is independent of the Hubble constant $h$, but contains the same amount of information on the growth of large scale structure.
In parallel with $\alpha_{||}$ and $\alpha_{\perp}$, we define
\begin{eqnarray}\label{eq:alphag}
    \alpha_{g}&\equiv& \frac{f_g(z) \sigma_m(z)}{f_{g,\mathrm{fid}}(z) \sigma_{m,\mathrm{fid}}(z)}
    =\frac{f_g(z) G(z) \sqrt{P_0}}
    {f_{g,\mathrm{fid}}(z)G_{\mathrm{fid}}(z)\sqrt{P_{0,,\mathrm{fid}}}}\nonumber\\
    &=&\frac{\beta(z) \sqrt{P_{\text{n}}(z)}}{f_{g,\mathrm{fid}}(z)G_{\mathrm{fid}}(z)\sqrt{P_{0,,\mathrm{fid}}}}, 
\end{eqnarray}
where $P_{n}(z)=P_0 G(z) b(z)$ is the normalization factor of the galaxy power spectrum at each redshift slice, with $b(z$) denoting the galaxy bias factor, and $\beta(z)=f_g(z)/b(z)$. Both $P_n(z)$ and $\beta(z)$ can be measured from observational data.

\subsection{Survey window function}
\label{sec:window}

Since the survey area in the simulation and actual observation is not regular, its effect on the measured galaxy power spectrum is not negligible. Applying this survey window function correction in configuration space is easy and straightforward, but becomes complicated in Fourier space. In our analysis, we adopt the correcting method for BOSS data (\citealt{Beutler_2017, Beutler_2017b}) as suggested by \cite{Wilson_2017}. 

The correction can be summarized as follows. We first calculate the power spectrum multipoles of a given model and Fourier transform them to get the correlation function multipoles $\xi_{\ell}$. Then we multiply $\xi_{\ell}$ with the window function to obtain the ``convolved" correlation function multipoles $\xi_{\ell}^{c}$. Finally we Fourier transform back $\xi_{\ell}^{c}$  to get the convolved power spectrum multipoles $P_{\ell}^{c}$ which can be compared with the measurements from galaxy mocks or observational data. The multipoles of correlation function and power spectrum are connected through Hankel transformation, i.e.
\begin{equation}
    P_{\ell}^{c}=4\pi(-i)^{\ell}\int dss^{2}\xi^{c}_{\ell}(s)j_{\ell}(sk),
\end{equation}
where $j_{\ell}$ is the spherical Bessel function of order $\ell$.

For actual observational data or galaxy mock as in this paper, we can calculate the multipoles of the window function from the random pair counts $RR(s,\mu)$ as (\citealt{Beutler_2017b})
\begin{equation}
    W^{2}_{\ell}(s)\propto RR(s,\mu)L_{\ell}(\mu),
\end{equation}
and the normalization is $W_{0}^{2}(s\to0)=1$. The application of the window function multipoles to the correlation function is given by
\begin{eqnarray}
    \xi_{0}^{c}&=&\xi_{0}W_{0}^{2}+\frac{1}{5}W_{2}^{2}+\mathrm{higher~order~term} \\
    \xi_{2}^{c}&=&\xi_{0}W_{2}^{2}+\xi_{2}[W_{0}^2+\frac{2}{7}W_{2}^{2}]+\mathrm{higher~order~term}
\end{eqnarray}
We are only interested in the monopole and quadrupole in this paper so we ignored the contribution from hexadecapole and higher order multipoles to the correction. In this work, we perform the Hankel transformation of the correlation function and power spectrum by using python package \textsf{Hankel} (\citealt{Murray2019}). In addition to the window function correction, we also consider the integral constraint bias as in \cite{Beutler_2017b}. For the galaxy mock in this paper, we find that this correction is only important at large scales. Since our following recovery analysis is limited to $k>0.02h$Mpc$^{-1}$, the results are not affected significantly by this cut.

\section{Data Analysis Results}\label{sec:BAO}

Given the mock catalog for the estimate of the covariance matrix and the galaxy power spectrum model in Sec.\ref{sec:pk}, we perform a MCMC analysis to derive the parameter constraints. We choose the monopole and quadrupole of the galaxy power spectrum in redshift space as the clustering statistics. We note that the observables in configuration space can provide complementary information, and higher order statistics can tighten the cosmological constraints; we leave those for future work. Since the number densities of galaxy at redshift $1.6<z<1.8$ and $1.8<z<2.0$ are significantly lower than the lower redshift counterparts, the measurement of the power spectrum is much noisier as in Figure \ref{fig:pk_ezmock}. Thus we combine these two redshift bins in the analysis below. 

\begin{figure*}
\begin{center}
\includegraphics[width=8.5cm]{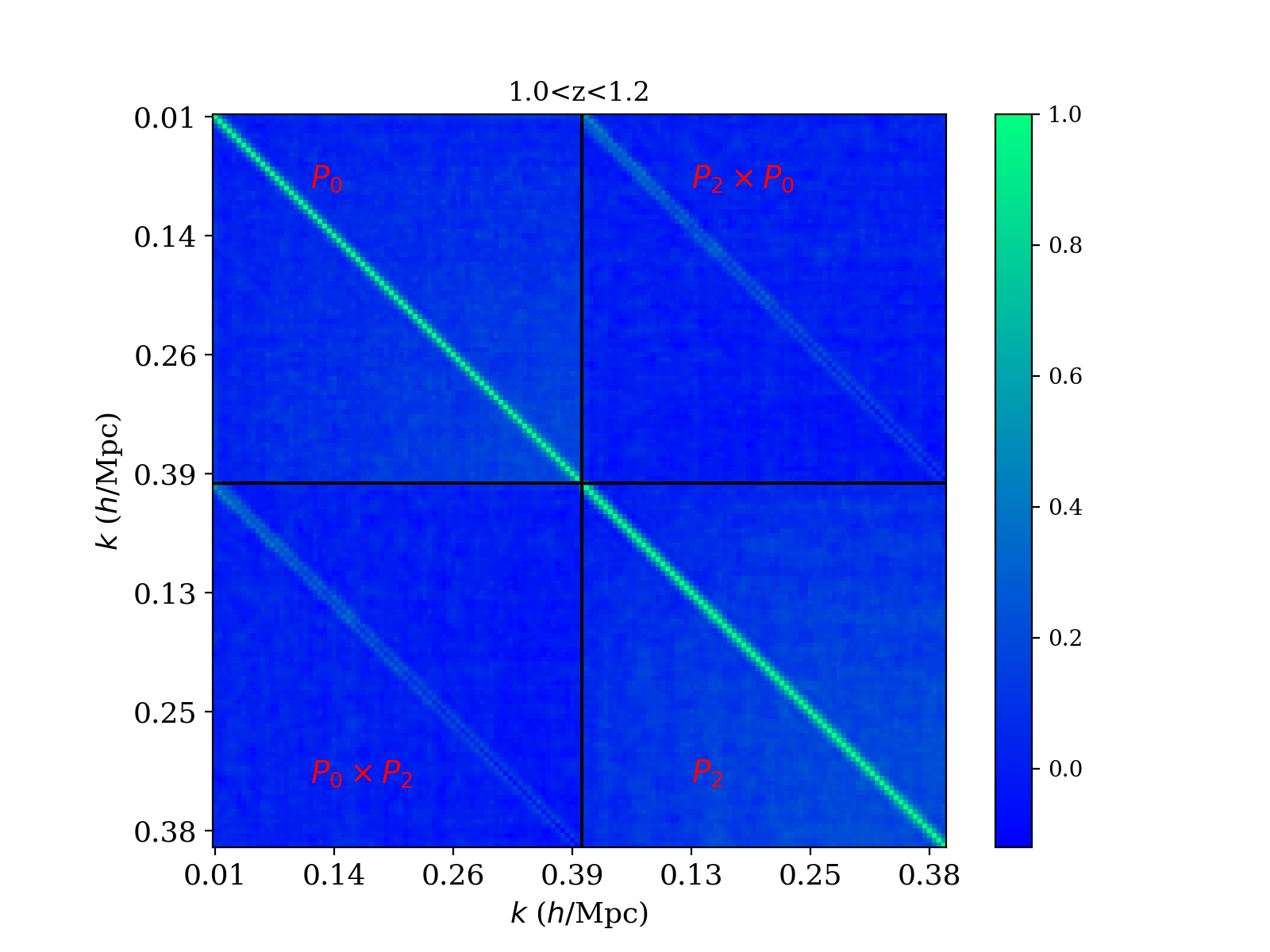}
\includegraphics[width=8.5cm]{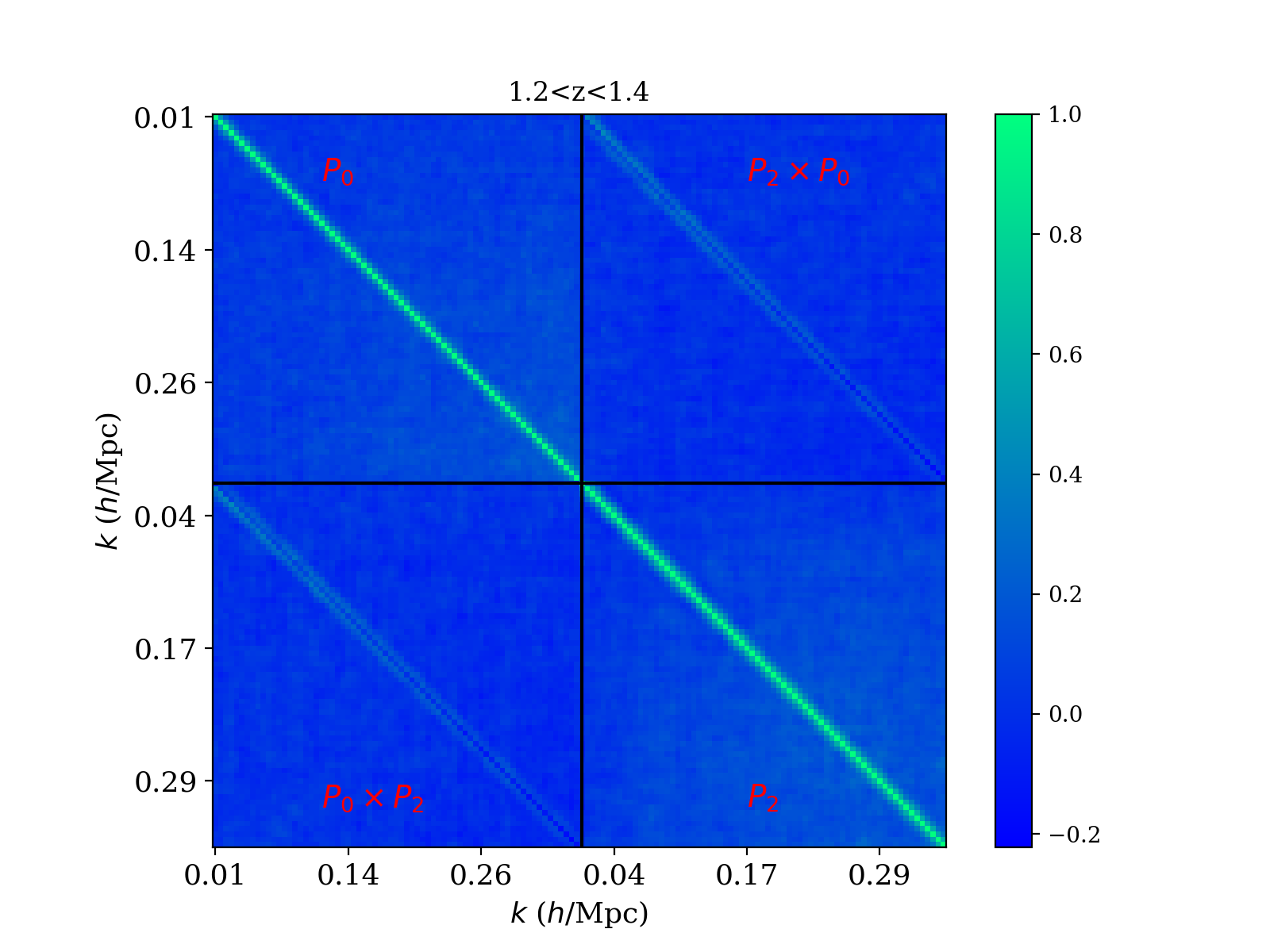}
\includegraphics[width=8.5cm]{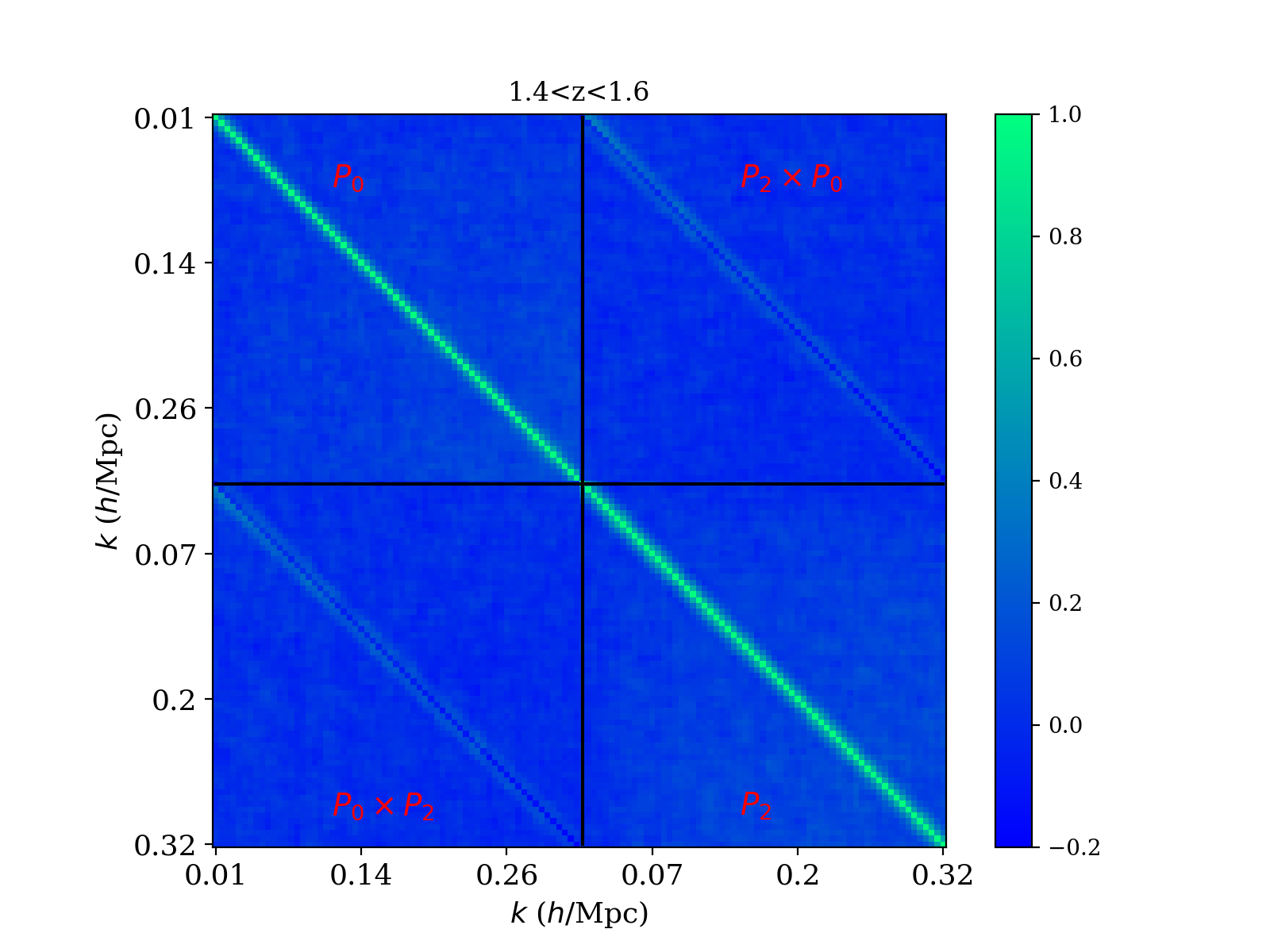}
\includegraphics[width=8.5cm]{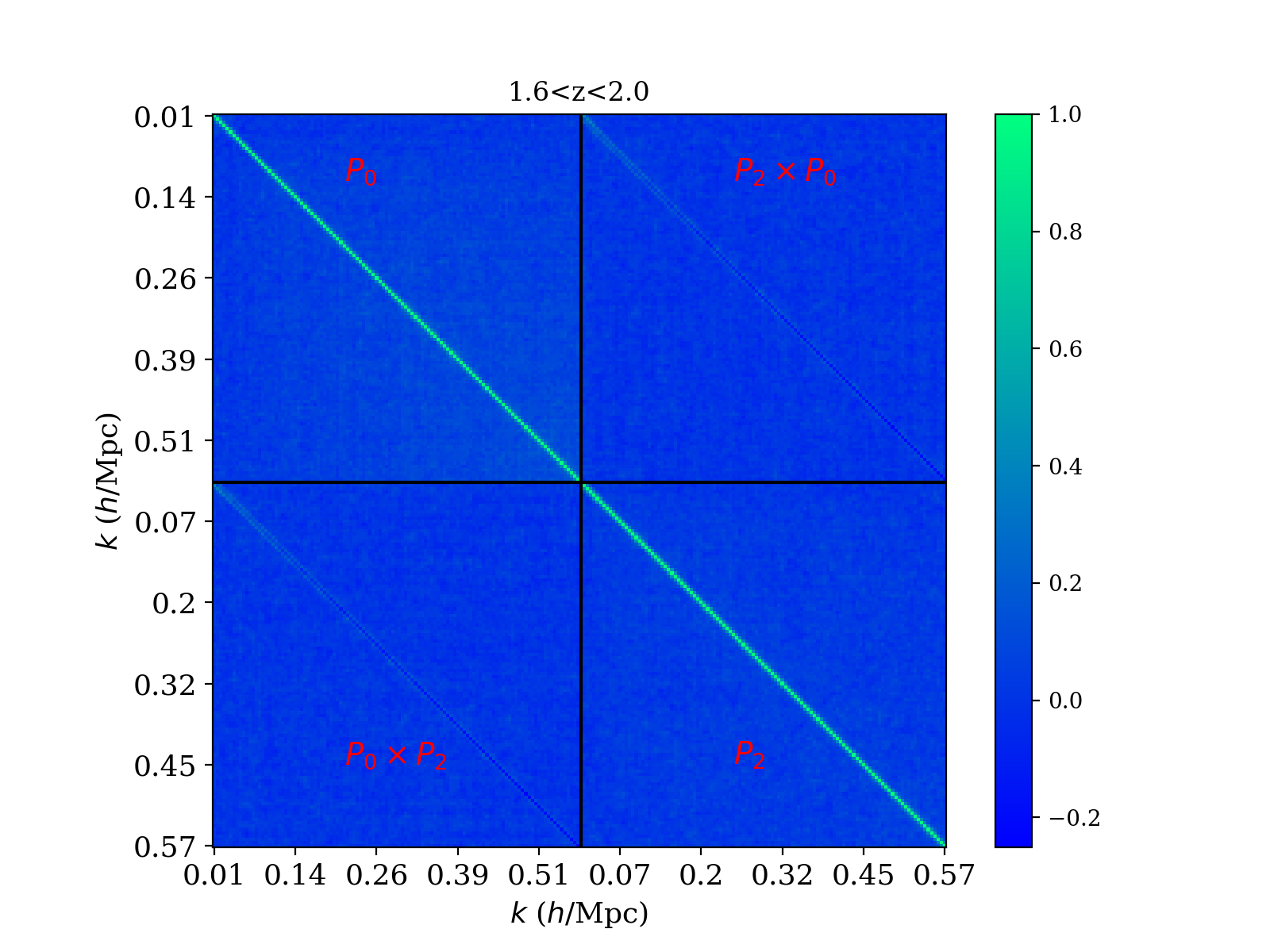}
\caption{Correlation matrix for monopole and quadrupole of the galaxy power spectrum in redshift space estimated from the EZmocks. The color indicates the level of correlation. In each figure, the correlation of monopole is shown in the upper-left corner, while the correlation for the quadrupole is in the lower-right panels. The off-diagonal corner shows the cross correlations. }
\label{fig:correlation_matrix}
\end{center}
\end{figure*}

In Figure \ref{fig:correlation_matrix}, we present the correlation matrix, i.e. normalized covariance matrix of monopole and quadrupole of the galaxy power spectrum in redshift space for each subsample. For each panel, the upper left corner shows the correlation between bins in monopole, the lower right corner presents correlation between bins in quadrupole, the lower left and upper right shows cross correlation between monopole and quadrupole. In redshift space we find that the $k$ modes are not independent, and this correlation can increase as we go to smaller scales.

\begin{figure*}
\begin{center}
\includegraphics[width=17.5cm]{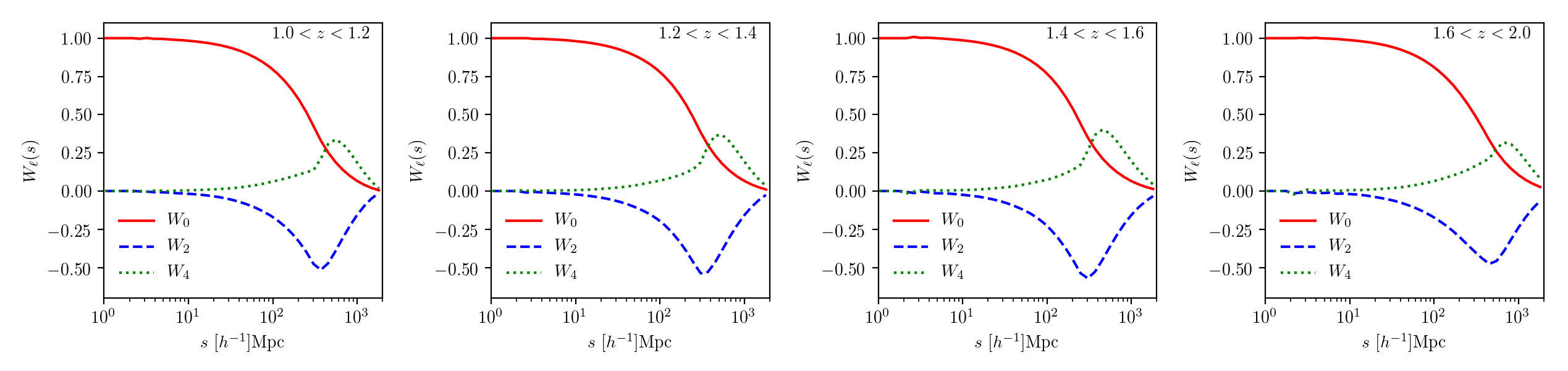}
\caption{Window function multipoles for the mock galaxy catalogs calculated through random pair counting. It shows the first few multipoles which are relevant to the analysis in this work. Each panel corresponds to a redshift slice as indicated in the upper right corner.}
\label{fig:window_Wl}
\end{center}
\end{figure*}

\begin{figure}
\begin{center}
\includegraphics[width=8.5cm]{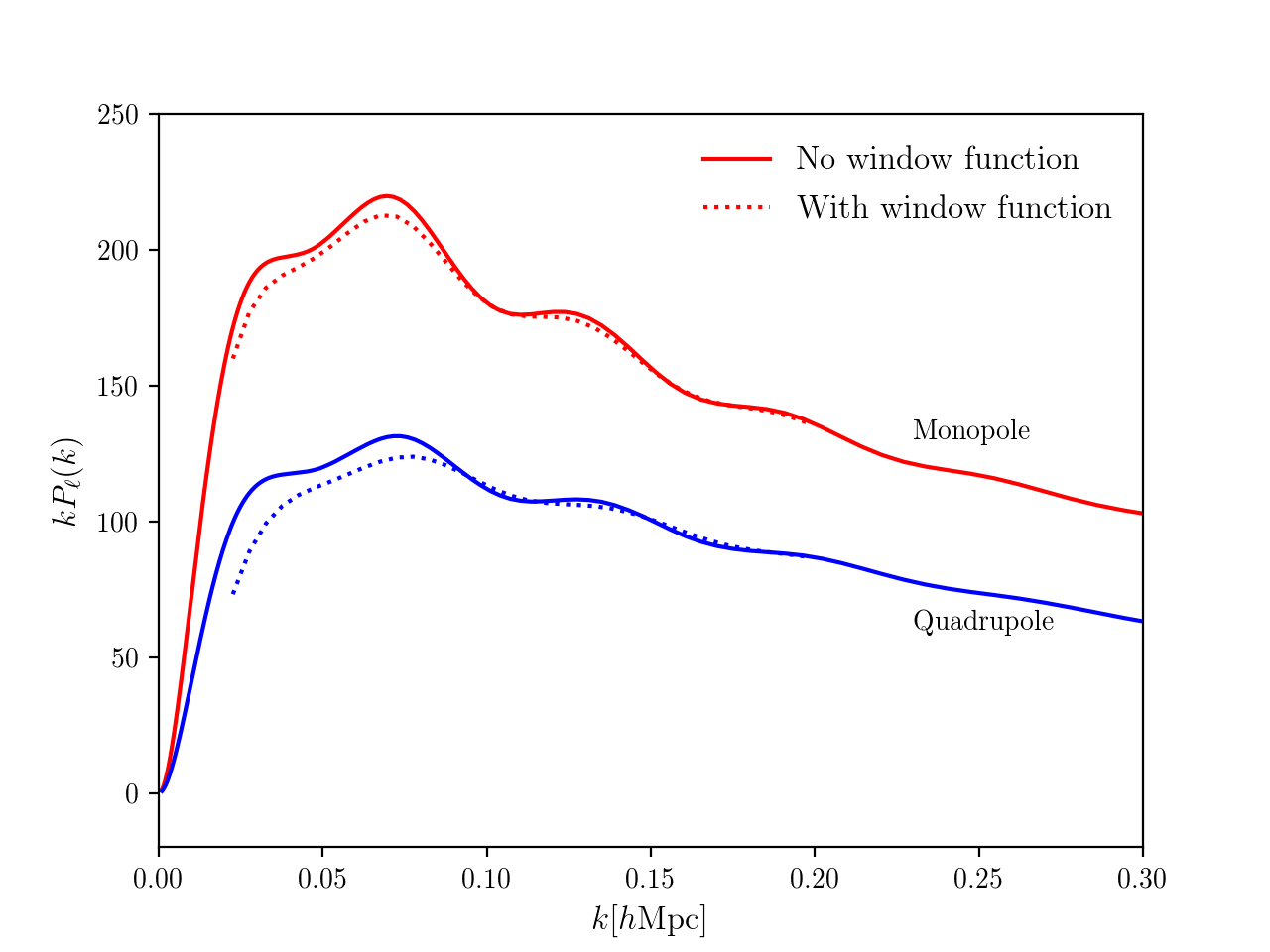}
\caption{Effect of the window function on the modeling of galaxy power spectrum, including monopole and quadrupole. The solid line is calculated from the power spectrum template as in Section \ref{sec:pk}, and the dotted line is after applying the window function correction as described in Section \ref{sec:window}.}
\label{fig:window_pk}
\end{center}
\end{figure}

In Figure \ref{fig:window_Wl}, we present the first few multipoles of the window function as described in the previous section. The result shows similar behavior as in the BOSS analysis (\citealt{Beutler_2017, Beutler_2017b}). Obviously, the effect of the window function on the galaxy power spectrum can be non-trivial and scale-dependent. As a sanity test, we present in Figure \ref{fig:window_pk} a comparison of the galaxy power spectrum with and without applying the window function correction. The result shows that the window function has significant impact at large scale and become minor at $k>0.1h$Mpc$^{-1}$, since the volume of the survey can limit the large scale modes. We note that without window function correction, the modeled galaxy power spectrum is higher than the truth (corrected), which means a model needs lower amplitude of power spectrum at linear scale to fit the data. This impact can lead to an underestimate of the relevant quantity, for instance the linear growth rate. We will present a relevant comparison later in this paper.

With the window function correction, we can compare the theoretical model with the observational data (here the simulated Roman HLSS H$\alpha$ mock galaxy catalog). We use Markov Chain Monte Carlo (MCMC) likelihood analysis in the recovery test. The likelihood function can be written as 
\begin{equation}
    \chi^{2}=\sum_{i,j}(P_{\text{obs, i}}-P_{\text{th,i}})C_{ij}^{-1}(P_{\text{obs,j}}-P_{\text{th,j}}),
\end{equation}
where $P_{\text{obs}}$ are the galaxy power spectrum measured from the reference Galacticus catalog, $P_{\text{th}}$ is the (window function corrected) prediction from the theoretical template as in Section \ref{sec:pk}, and $C$ is the covariance matrix estimated from EZmocks. Since the estimated covariance matrix is from mock catalogs, the inverse $C^{-1}$ is biased compared with the true inverse covariance matrix (\citealt{Hartlap_2007}). Thus we need to correct for this bias by rescaling the inverse covariance matrix as
\begin{equation}
    C^{-1}_{\text{Hartlap}} = \frac{N_{s}-n_{b}-2}{N_{s}-1}C^{-1},
\end{equation}
where $N_{s}=1000$ is the number of mocks used in the estimate of the covariance matrix, and $n_{b}$ is the number of bins of galaxy power spectrum used in the likelihood analysis. For the galaxy samples analyzed in this work, this correction can increase the final uncertainty by a few percent.

We use the python package emcee \citep{ForemanMackey_2013} to run the MCMC test and get the posterior of the parameters. In our method, the parameter set for the galaxy power spectrum is \{$\alpha_{||}, \alpha_{\perp}, k_{*}, P_{\text{n}}, \beta, \sigma_{v}, A, Q, B, f_{g}^*$\}. Note that $f_{g}^*$ is only used in Eq.[\ref{eq:g_mu}], and not used in measuring the growth rate $f_g$.
We use uninformative and flat priors on these parameter throughout this work: $\alpha_{||}=[0.2, 1.8]$, $\alpha_{\perp}=[0.2, 1.8]$, $k_{*}=[0.01, 0.8]h$Mpc$^{-1}$, $
\log{P_{\text{n}}}=[2, 10]$, $\beta=[0.01, 2.0]$, $\sigma_{v}=[0.1, 1000]$km s$^{-1}$, $A=[0.5, 15]$, $Q=[2.0, 30.0]$, $B=[0.2, 10.0]$ and $f_{g}^*=[0.01, 2.0]$. For Model B, we have an additional parameter $\Delta\alpha(z)$ with prior $[0,2]$

\begin{figure*}
\begin{center}
\includegraphics[width=8.5cm]{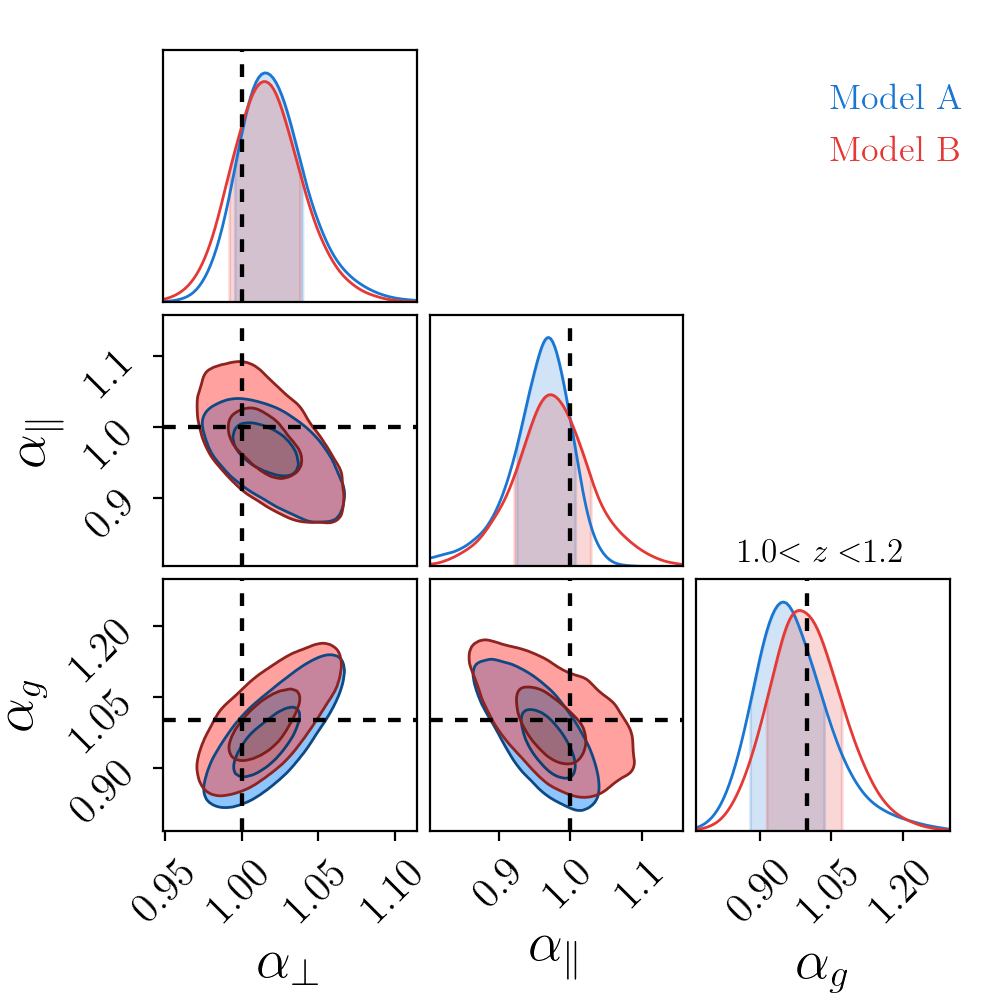}
\includegraphics[width=8.5cm]{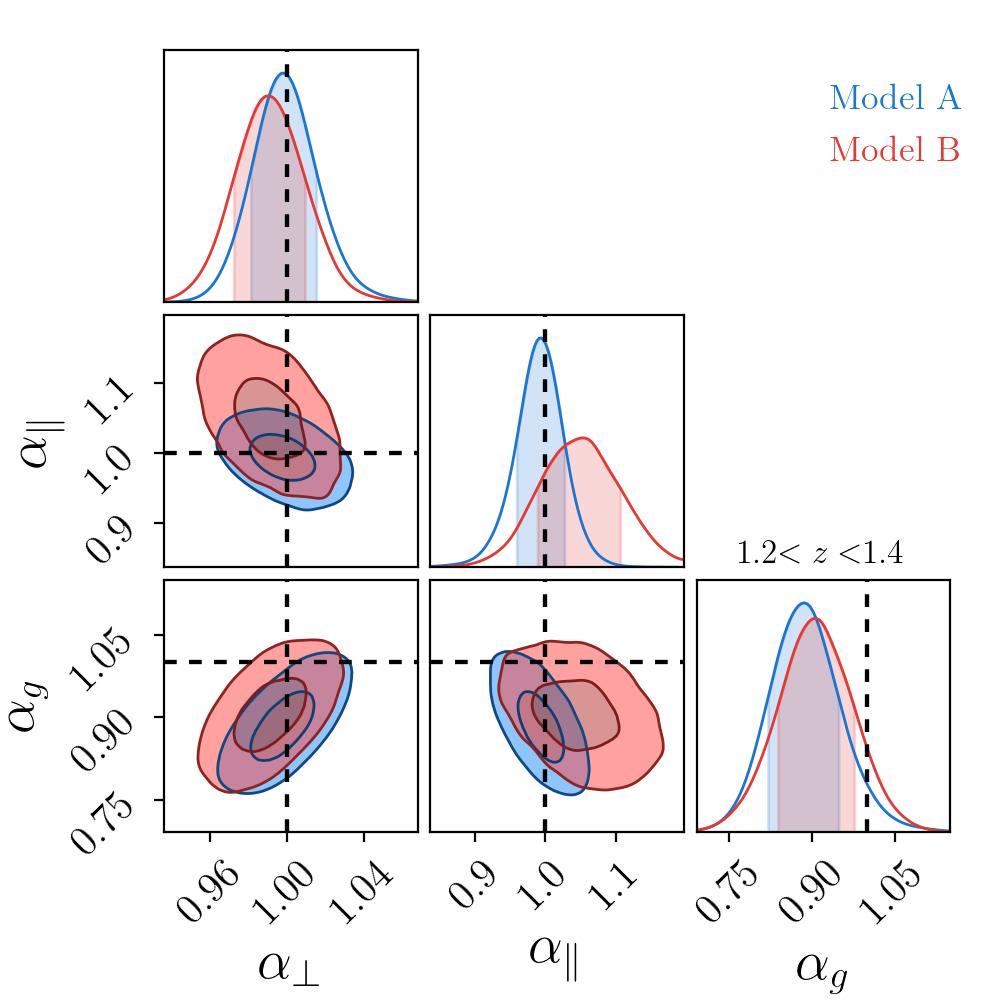}
\includegraphics[width=8.5cm]{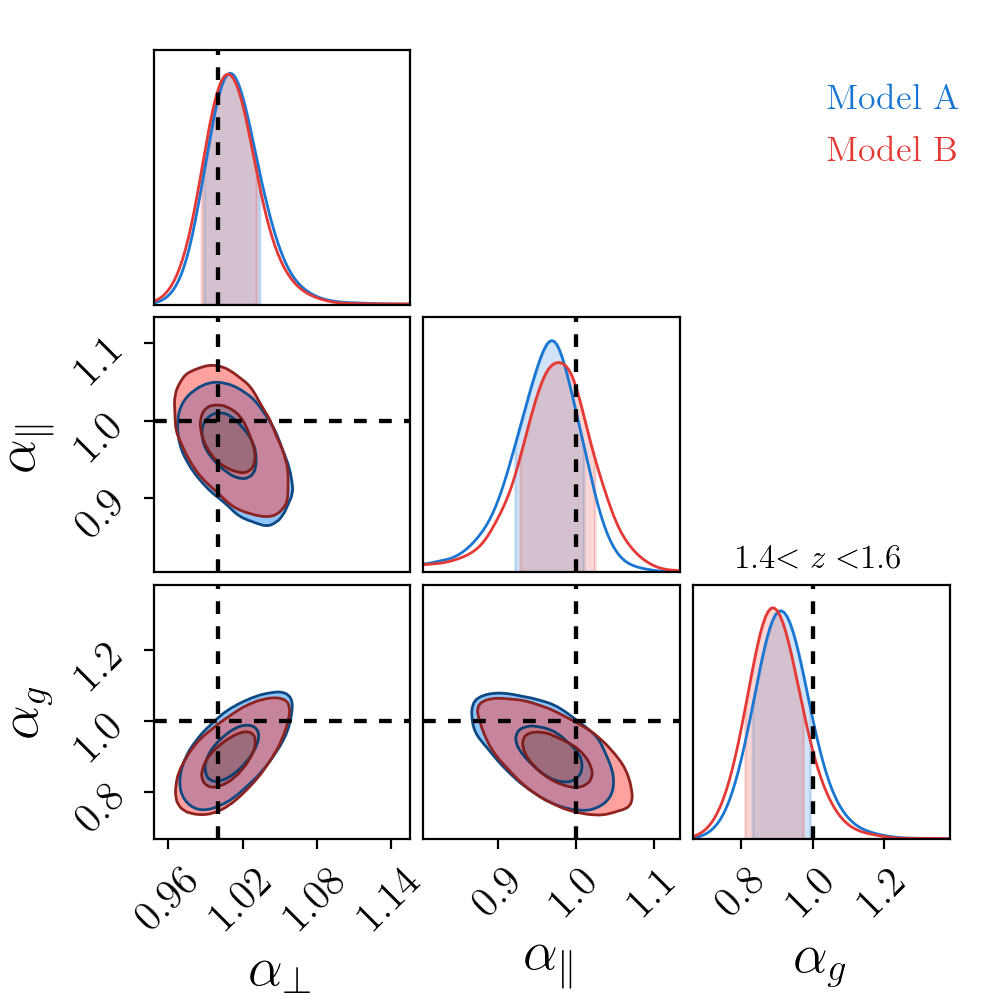}
\includegraphics[width=8.5cm]{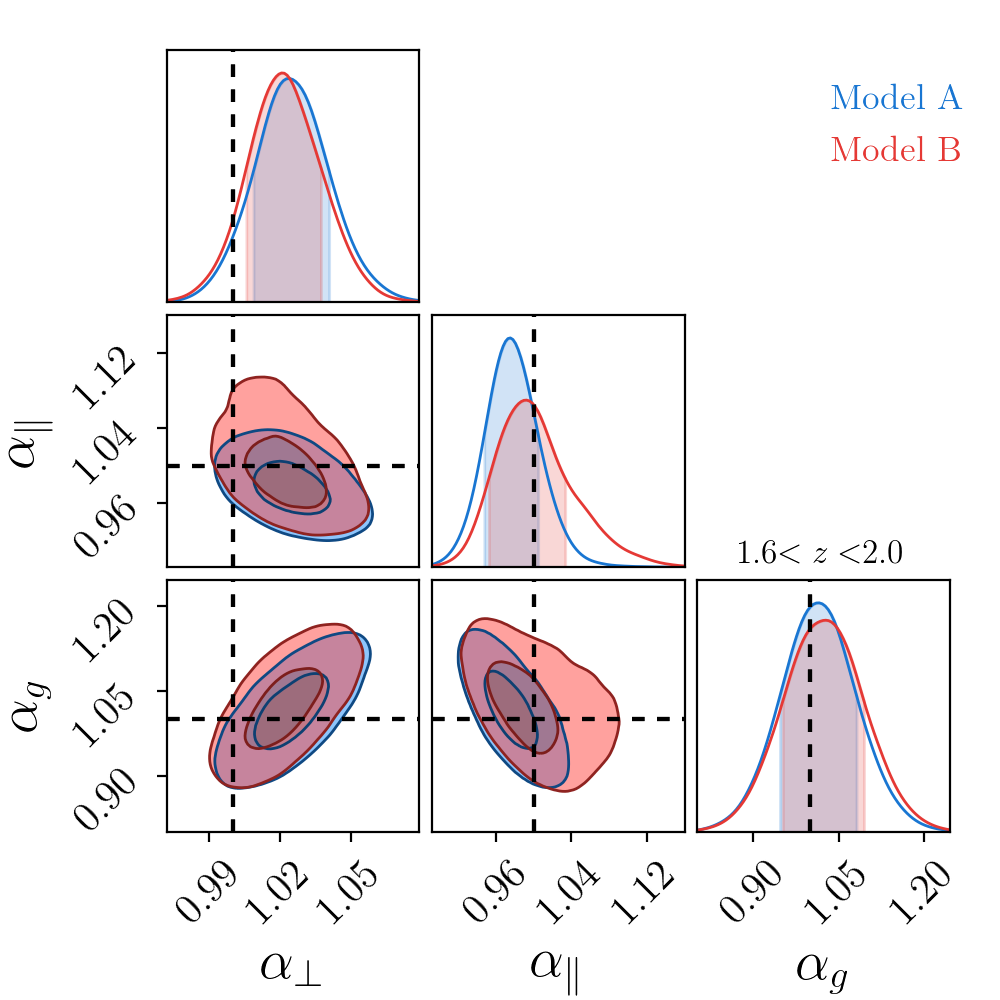}
\caption{2D and 1D distribution of the scaling parameters $\alpha_{||}$, $\alpha_{\perp}$ and $\alpha_{g}$ from the MCMC analysis, marginalized over other parameters. We show contours at 68 and 95 percent confidence levels with $k_{\text{max}}=0.3$ for two different templates of galaxy power spectrum. The dashed lines represent no deviation from the fiducial model $\alpha_{||}=\alpha_{\perp}=\alpha_{g}=1$. The overall result validates the simulation and modeling of galaxy clustering from the Galacticus catalog. The fiducial cosmology can be recovered at $1-2\sigma$ level for all the subsamples.}
\label{fig:BAO_constraint}
\end{center}
\end{figure*}

We restrict the use of the galaxy power spectrum at $0.02<k<0.3~h$Mpc$^{-1}$ in the analysis, indicating that the information at small scale is also used. In Figure \ref{fig:BAO_constraint}, we present the constraints on the scaling parameters $\alpha_{||}$, $\alpha_{\perp}$ and $\alpha_{g}$ after marginalizing over other parameters. The fiducial value of these parameters are recovered at 1 to 2$\sigma$ level which validates our construction of the galaxy mock and theoretical modeling. The results show that the two different models of the galaxy power spectrum can give parameter constraints that are consistent within 1$\sigma$. For Model B, the addition of parameter $\Delta\alpha(z)$ increases the measurement uncertainties somewhat, but leads to more accurate constraints on parameters $\alpha_{||}$ and $\alpha_{g}$ in most cases, both effects are as expected.

\begin{figure}
\begin{center}
\includegraphics[width=8.5cm]{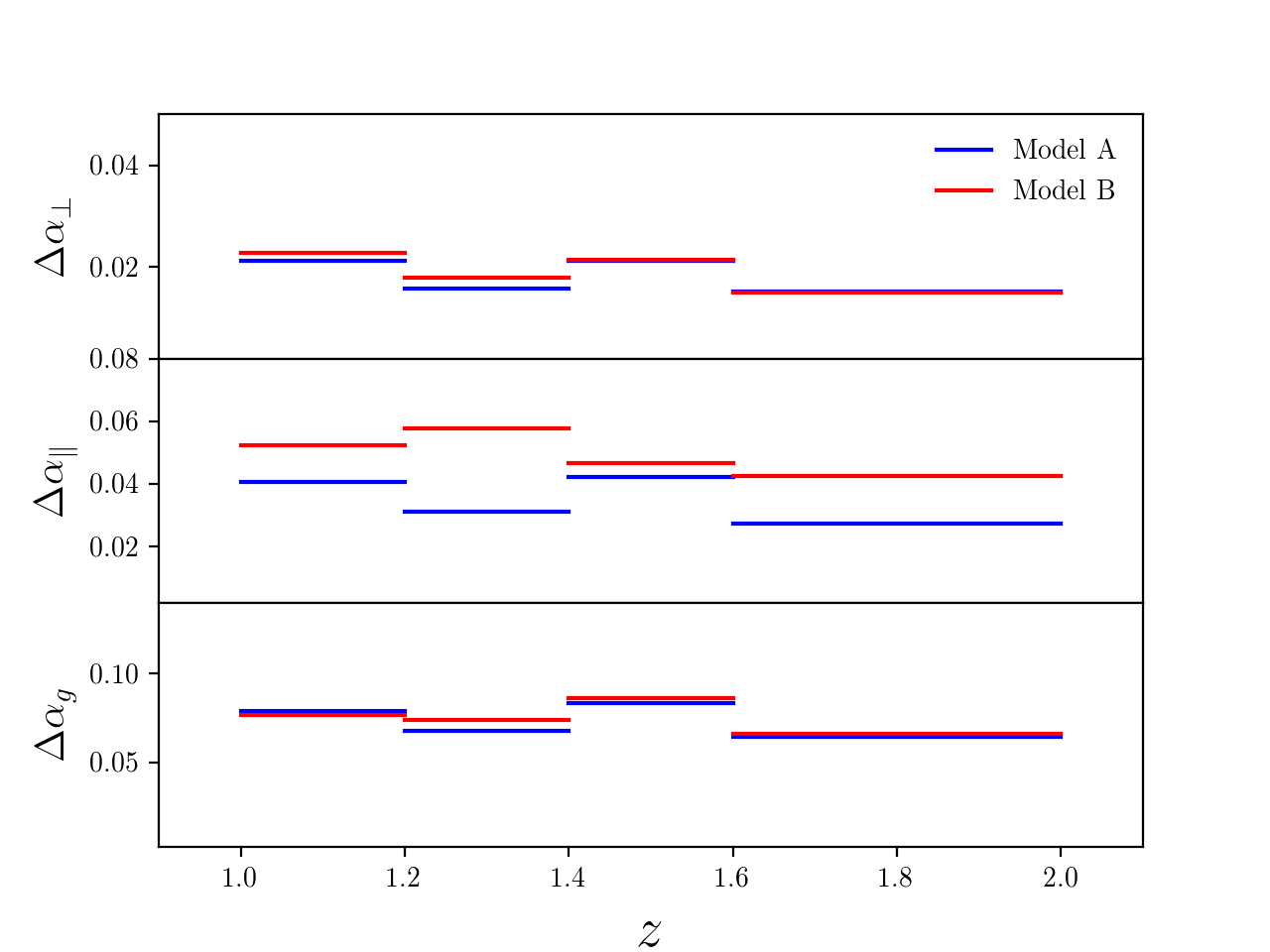}
\caption{Uncertainties of the scaling parameters measured from H$\alpha$ emitting galaxies by RST, after projection from the constraint in previous section. We display estimates of both templates of galaxy power spectrum.}
\label{fig:BAO_error}
\end{center}
\end{figure}

Figure \ref{fig:BAO_error} shows the 1D uncertainties of $\alpha_{||}$, $\alpha_{\perp}$ and $\alpha_{g}$ expected from the Roman HLSS H$\alpha$ galaxies. Within redshift range $1<z<2$, we find that the clustering analysis can give 2\% measurement of $\alpha_{\perp}$, i.e. the angular diameter distance. For $\alpha_{||}$, i.e. the Hubble parameter, the measurement is accurate at 3-6\% level depending slightly on the analysis template. For $\alpha_{||}$ (see Eq.[\ref{eq:alphag}]), i.e., the linear growth parameter $f_g\sigma_{m}$ (see Eq.[\ref{eq:fgsigmam}]), the measurement is at roughly 7\%. Compared with the current measurements, Roman can give accurate measurement of these dark energy observables using a single tracer over a wide redshift coverage. 

\begin{figure}
\begin{center}
\includegraphics[width=8.5cm]{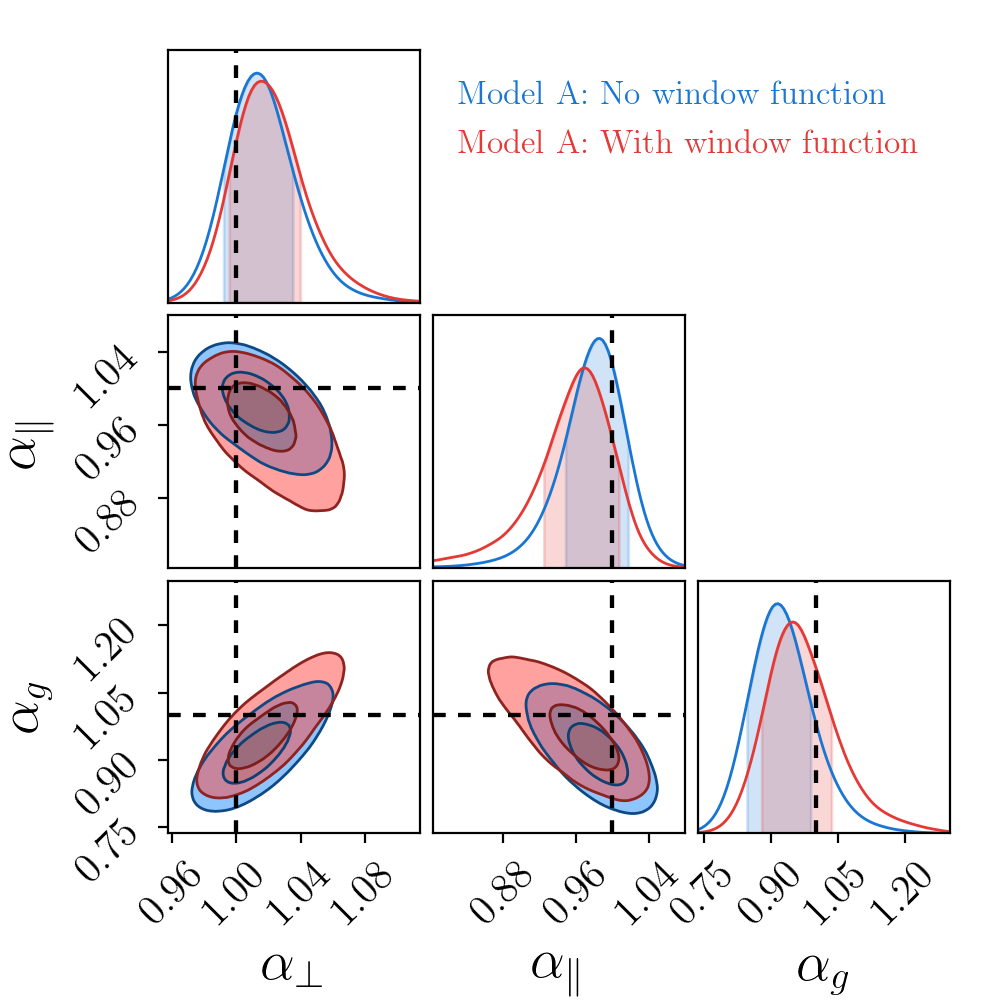}
\caption{Comparison of the constraint on the parameters $\alpha_{||}$, $\alpha_{\perp}$ and $\alpha_{g}$ with and without window function correction. This illustration uses galaxies with $1.0<z<1.2$ and model A for the template of galaxy power spectrum.}
\label{fig:window_constraint}
\end{center}
\end{figure}

In Figure \ref{fig:window_constraint}, we investigate the effect of window function on the cosmological parameter constraint. As an example, we present the result from galaxies with $1.0<z<1.2$ with model A for the galaxy power spectrum. It shows that the window function has significant impact on the estimate of radial distance scale and linear growth rate. For the former, the reason is that the maximum distance scale in the line-of-sight direction is smaller due to the thickness of the redshift shell which means the window function starts to work at smaller scales than the angular direction. This leads to an overestimate of the Hubble parameter at percent level. The effect on the linear growth rate can be seen from Figure \ref{fig:window_pk} as well. Without window function correction, the model galaxy power spectrum can have a lower amplitude. According to Eq (\ref{eq:alphag}), this leads to an underestimate of the linear growth rate. For galaxy mock in this paper, this change can be as large as 5\%. Although the effect of the window function on all the parameters is smaller than 1$\sigma$, this is non-negligible for precision cosmology. In addition, the analysis in this paper uses information up to $k=0.3h$Mpc$^{-1}$, the window function effect is reduced to some extent by incorporating this non-linear information. Therefore it is likely that a smaller $k_{\mathrm{max}}$ can lead to more biased result.

\section{Discussion and Conclusion}

Clustering of galaxies provides important information of the evolution of the universe, and a powerful probe of dark energy. In particular, galaxy clustering measured from galaxy redshift surveys enables measurement of cosmic expansion history and linear growth rate as functions of redshift, which can be used to determine the true nature of the observed cosmic acceleration. For next generation surveys like those planned for Roman and Euclid, the observation of millions of galaxies can enable accurate measurements of the galaxy clustering signal, and the resulting cosmological measurements can help to constrain the dark energy and modified gravity models, or any deviation from standard $\Lambda$CDM model. In this paper, we use Galacticus, a calibrated SAM to generate a realistic synthetic catalog of H$\alpha$ emission line galaxies that Roman will observe in its HLSS. This catalog has the same area on the sky, redshift range, and flux limit as specified by the current mission baseline design, and can serve as a reasonable reference to evaluate the expected performance. 
In particular, we have applied the calibrated SAM Galacticus of galaxy formation evolution from \cite{Zhai_2019MNRAS} to the dark matter merger trees from the UNIT simulation described in \citet{Chuang_2019} %{\citealt{Chuang_2019}},
and built a lightcone catalog of emission line galaxies within redshift $1<z<2$. After applying a dust attenuation model, we select all the galaxies with H$\alpha$ flux higher than $1\times10^{-16}$erg s$^{-1}$ cm$^{-2}$ to form our final galaxy sample. 

We have measured the clustering statistics of this mock catalog, including 2PCF and power spectrum. In order to validate our algorithm and forecast the dark energy measurements from Roman HLSS, we use EZmock code to produce thousands of approximate but fast galaxy mocks to estimate the covariance matrix for the statistics of interest. These EZmocks are calibrated to match the clustering of the reference SAM catalog. In particular, we employ an emulator approach to search the parameter space of EZmock and the resulting model can fit the monopole of 2PCF and power spectrum in both real and redshift space. The calibrated EZmocks are further truncated to have the same angular and radial distribution as the reference model, which makes the estimate of the covariance matrix straightforward. We note that the calibration of EZmock is not perfect given the uncertainty in the clustering measurement, this can be further improved if higher order statistics are taken into account. 

With these mock galaxy catalog, we perform a recovery test using the clustering signal in Fourier space. In particular, we compare two RSD models of the observed galaxy power spectrum and perform a likelihood analysis to get the posterior of the parameters. The result shows that Roman galaxy redshift survey can measure the angular diameter distance at close or better than 2\%, the Hubble parameter can be measured at 3-6\% accuracy, and the error of the linear growth parameter is at 7\%. 
The comparison of the results form these two different RSD models indicate that RSD modeling has a significant impact on the accuracy and precision of the dark energy measurements.

Our analysis has been performed using power spectrum due to the simplicity in calculating its theoretical prediction. It is also possible to analyze in configuration space, which can provide a consistency check on the cosmological constraints. 

We have included the window function correction in the interpretation of the galaxy power spectrum. We find that not accounting for the window function correction leads to the under-estimate of parameter errors, and a biased estimate of the linear growth parameter.

The galaxy mock constructed in this work can be used beyond the two point statistics analysis presented here. The constraining power on cosmological models from Roman H$\alpha$ emission line galaxies can be further explored by higher order statistics, cross correlation with other dark matter tracers, galaxy-galaxy lensing, statistical properties of the void distribution in the universe and so on. Our mock catalog enables the investigation of these topics for the Roman HLSS data. Such analysis will greatly strengthen the dark energy constraining power of Roman HLSS.

Note that our recovery test on the galaxy mock has been performed to measure the BAO and RSD signals. The actual measurements from Roman can be impacted by multiple factors. The dust attenuation can change the number and density of galaxies we can observe, which in turn can change the uncertainty of the clustering measurement, as well as reducing or increasing the shot noise, the survey volume may have an effect on the scales that we can probe and use, contamination from other emission lines such as [NII] may change the identification of H$\alpha$ emission line galaxies and affect the bias of the galaxy sample, the actual survey strategy may be different from the assumed in this paper, a shallower or deeper flux sensitivity can change the observation of faint galaxies, and the resulting galaxy sample due to these low mass halos is also affected. While a comprehensive exploration of these systematic effects on the cosmological constraint from Roman galaxy redshift survey is beyond the scope of this paper, the Roman galaxy mock we have presented here enables realistic studies of cosmological constraints that can be expected from Roman HLSS. 

We note that for the purpose of forecasting the cosmological constraints from a galaxy survey, the simple Fisher matrix formalism can be used to investigate the impact of changing the galaxy density, survey volume, scales of clustering in the analysis, and different types of dark energy properties and modified gravity theories. However, this method usually gives a lower limit of the uncertainties on the parameters of interest, which means the forecast for the cosmological parameters are more idealistic than the simulation based method as we applied in this paper. Our Roman HLSS H$\alpha$ galaxy mock can be used in carrying out sanity checks for Fisher matrix based studies. Finally, the pipeline for the mock construction and data analysis presented in this work is directly applicable to the analysis of real observational data, thus it lays a solid foundation for probing dark energy using future spectroscopic surveys. 

% End of mnras_template.tex

\section*{Data Availability}
The original dark matter halo catalogs are available from the UNIT simulation website. The galaxy mocks are available by request. A public webpage presenting the mocks will be available at a later time.

\section*{Acknowledgements}

This work is supported in part by NASA grant 15-WFIRST15-0008, Cosmology with the High Latitude Survey Roman Science Investigation Team (SIT). GY would like to thank MICIU/FEDER (Spain)  for financial support under project grant  PGC2018-094975-B-C21.
The UNIT simulations have been done in the MareNostrum Supercomputer at the Barcelona Supercomputing Center (Spain) thanks to the  cpu time awarded by PRACE under project grant number 2016163937. This work used the Extreme Science and Engineering Discovery Environment (XSEDE), which is supported by National Science Foundation grant number ACI-1548562 (\citealt{XSEDE_2014})

\rm{Software:} Python,
Matplotlib \citep{matplotlib},
NumPy \citep{numpy},
SciPy \citep{scipy}, George \citep{george_2014}, emcee \citep{Foreman-Mackey_2013}, Hankel \citep{Murray2019}.

%\software{Python,
%Matplotlib \citep{matplotlib},
%NumPy \citep{numpy},
%SciPy \citep{scipy}
%}

\bibliographystyle{mnras}
\bibliography{emu_gc_bib,software}

\bsp	% typesetting comment
\label{lastpage}
\end{document}